\documentclass[sigconf,nonacm]{acmart}

\thanks{This is an extended version of the paper published in \textit{ASIA CCS '26: Proceedings of the ACM Asia Conference on Computer and Communications Security}~\cite{DSilva26-CR}.}

\usepackage{multirow}
\usepackage{svg}
\usepackage{siunitx}
\usepackage{graphicx}
\usepackage{makecell}
\usepackage{pifont}
\usepackage{enumitem}
\usepackage{mathtools}
\usepackage{wasysym}
\usepackage{stmaryrd}
\newcommand{\COMMENTBLOCK}[1]{}
\newcommand{\solidcircle}{\CIRCLE}
\newcommand{\emptycircle}{\Circle}
\newcommand{\halfcircle}{\LEFTcircle}

\begin{document}
\title{SoK: Practical Aspects of Releasing Differentially Private Graphs}

\author{Nicholas D'Silva}
\orcid{0009-0008-6195-7662}
\affiliation{%
  \institution{The University of New South Wales}
  \city{Sydney}
  \state{NSW}
  \country{Australia}}
\email{n.dsilva@unsw.edu.au}

\author{Surya Nepal}
\orcid{0000-0002-3289-6599}
\affiliation{%
  \institution{CSIRO's Data61 \and The University of New South Wales}
  \city{Sydney}
  \state{NSW}
  \country{Australia}}
\email{surya.nepal@data61.csiro.au}

\author{Salil S. Kanhere}
\orcid{0000-0002-1835-3475}
\affiliation{%
  \institution{The University of New South Wales}
  \city{Sydney}
  \state{NSW}
  \country{Australia}}
\email{salil.kanhere@unsw.edu.au}

\renewcommand{\shortauthors}{D'Silva et al.}

\begin{abstract}
Graph data is increasingly prevalent across domains, offering analytical value but raising significant privacy concerns. Edges may encode sensitive relationships, while node attributes may contain sensitive entity or personal data. Differential Privacy (DP) has gained traction for its strong guarantees, yet applying DP to graphs is challenging because of their complex relational structure, leading to trade-offs between privacy and utility. Existing methods vary in privacy definitions, utility goals, and contextual settings, complicating comparison. For practitioners, this is compounded by DP's interpretability issues, contributing to misleading protection claims.

To address this, we propose a novel systemisation of existing methods tailored to practical considerations and adaptable to varying practitioner objectives. Our contributions include: (i) a comprehensive survey of differentially private graph release methods; (ii) identification of key vulnerabilities; and (iii) a practitioner-oriented, objective-based framework to guide the selection, interpretation, and sound evaluation of existing methods. We demonstrate the use of our systemisation through two exemplary scenarios in which we assume the role of a social network analyst, apply it, and conduct evaluations in accordance with our framework. Together, these two illustrative instantiations ultimately provide a unified benchmark for state-of-the-art methods in the social networks domain.
\end{abstract}




\maketitle

\section{Introduction}
Graphs are ubiquitous, foundational data structures, valued for their ability to model complex relationships between entities. They arise across diverse domains, including social networks, financial markets, transport networks, and biological processes~\cite{Barabasi16, Bonifati20}. Advances in data collection technologies, including smartphones, smart sensors, and other tracking-enabled devices, have significantly increased the scale and richness of graph-structured data. This growth has been accelerated by the adoption of graph-based systems, including social media platforms, Internet of Things networks, and graph databases~\cite{snap, ogb, nr}. These developments underscore the need for analytical methods that extract meaningful, domain-relevant insights from graphs.

The increased collection and analysis of graph data raises significant privacy concerns~\cite{Kifer11, Yuan24}. While such data offers high analytical value, both structural (e.g., edges representing relationships) and attribute information (e.g., demographic or behavioural details) can be exploited to compromise privacy~\cite{Yuan24}. Risks intensify when edges encode sensitive relationships, nodes represent entities requiring privacy, or attributes represent sensitive personal information. Privacy breaches may arise intrinsically from the data (e.g., private associations represented by edges~\cite{Korolova08}, or private community memberships~\cite{Ji19}), or through linkage with external sources that enable inferences about individuals' beliefs, affiliations, or other sensitive characteristics~\cite{Narayanan09}. The rise of graph neural networks (GNNs) has amplified these risks, as tasks such as community detection, link prediction, and node classification or re-identification can inherently expose private information~\cite{Zhang24a}. Protecting graph data while preserving its utility thus remains a critical challenge, often framed as the \textit{privacy-utility trade-off}.

Differential Privacy~\cite{Dwork06} (DP) has emerged as a leading privacy framework, offering formal guarantees through calibrated noise addition to algorithm outputs, ensuring minimal impact from any single record. While effective, this noise degrades the dataset's utility, reinforcing the fundamental privacy-utility trade-off~\cite{Dwork06}. This challenge is even greater for graphs, which have a more complex relational structure than tabular data.

Several methods for private graph release have been proposed~\cite{Yuan23, Yang21, Chen20, Qin17}, often optimising utility for specific applications. However, the diversity of utility definitions (e.g., preserving certain structural features) and evaluation strategies makes meaningful cross-method comparison difficult.

Several surveys consolidate methodologies and their privacy guarantees~\cite{Liu24, Li23, Mueller22, Casas-Roma17s}. However, practical considerations such as utility, vulnerabilities, and validity in practice are often limited or fragmented due to diverse perspectives. Which utility metrics matter most? Which methods align with our threat model? Are the assumptions made by some privacy mechanisms acceptable? Practitioners often seek answers to these questions; however, to our knowledge, no work provides clear guidance, leaving these considerations ambiguous and challenging to navigate. This hinders the adoption of DP graph mechanisms in real-world settings. We posit that this challenge will persist unless practitioner objectives are scoped to their specific contexts, especially when the ultimate goal is private data release. Such trade-offs are highlighted in~\cite{Stadler22}. 

The usability and interpretability of DP guarantees remain a challenge~\cite{Ivoline24}, especially for graphs where structural correlations add complexity. Practitioners, including non-experts, must understand the validity and assumptions of existing methods. This underscores the need for a flexible framework that increases the understandability of DP-based graph release approaches and adapts to evolving practitioner objectives regarding both privacy and utility.

\paragraph{Contributions} This work addresses these issues through several key advances. First, we conduct an updated, comprehensive survey of existing methods and present a novel layer-based systemisation whose modular design clarifies methodological components and intended purposes. We also add a dimension that exposes vulnerabilities arising from specific design choices, an aspect that prior work has mainly overlooked. Next, we propose a practitioner-oriented, objective-based framework to guide practitioners and non-specialists in selecting, applying, and evaluating privacy mechanisms tailored to their contexts and requirements, thereby improving interpretability and mitigating critical risks of invalid application or misleading data protection claims. To demonstrate our systematisation, we construct two exemplar scenarios in which we assume the role of a social network analyst, apply our framework, and conduct empirical evaluations,  providing a comprehensive benchmark of state-of-the-art (SotA) methods for social network analysis. To summarise, our contributions are:
\begin{enumerate}[leftmargin=*]
    \item We propose a novel systemisation of differentially private graph release methods from practical perspectives, consisting of: (i) a modular systemisation of approaches; (ii) an additional dimension addressing vulnerabilities and potential privacy attacks; and (iii) a practitioner-oriented framework to guide method selection, application, and evaluation for specific contexts;
    \item We construct two illustrative scenarios (assuming the roles of social network analysts), demonstrating the framework in action and providing empirical evaluations;
    \item A benchmark of SotA methods for social networks; and
    \item We release all our source code and results as open-source\footnote{\url{https://github.com/ndsi6382/SoK-PAoRDPG}\label{fn:repo}}, to support future research.
\end{enumerate}
To the best of our knowledge, no existing survey, framework, or systematisation addresses these critical practical considerations in DP-based graph release methods. The exact scope of the approaches we consider is detailed in Sec.~\ref{systemisation}.

\paragraph{Organisation}
We discuss related and foundational work in Sec.~\ref{related-work}. We present our systemisation in Sec.~\ref{systemisation}, and provide exemplar instances of our systemisation, including empirical evaluations, in Sec.~\ref{evaluations}. Future directions and limitations are discussed in Sec.~\ref{limitations-future-work}, with conclusions in Sec.~\ref{conclusion}. We provide further details in the \hyperref[formal-definitions]{Appendix}.

\section{Related Work}\label{related-work}
A growing body of research addresses DP-based graph release. We identify four key areas that warrant discussion for related work:
\begin{enumerate}[leftmargin=*]
    \item \textit{Literature surveys} that establish formal theoretical frameworks, definitions, and taxonomies for this domain;
    \item \textit{Empirical evaluations} that assess real-world utility, practicality, and applicability of SotA methods;
    \item Coverage of contemporary \textit{differential privacy-preserving techniques} for dataset publication in general, encompassing perturbative and generative approaches; and
    \item Coverage of \textit{graph-specific methods} designed to tackle challenges unique to graph-structured data, which differ substantially from those in tabular data.
\end{enumerate}
A summary of related work is presented in Table~\ref{tab:related-work}.

\begin{table}
    \small
    \centering
    \begin{tabular}{c|cccc}
        \hline \makecell{\textbf{Work}} & \makecell{\textbf{Literature}\\\textbf{Survey}} & \makecell{\textbf{Empirical}\\\textbf{Evaluation}} & \makecell{\textbf{Privacy}\\\textbf{Coverage}} & \makecell{\textbf{Graph}\\\textbf{Coverage}} \\
        \hline This Work & \solidcircle & \solidcircle & \solidcircle & \solidcircle \\
        \hline \cite{Chen25} & \emptycircle & \solidcircle & \solidcircle & \emptycircle \\
        \hline \cite{Hu24} & \solidcircle & \emptycircle & \solidcircle & \emptycircle \\
        \hline \cite{Liu24} & \emptycircle & \halfcircle & \solidcircle & \solidcircle \\
        \hline \cite{Guo23} & \solidcircle & \emptycircle & \emptycircle & \solidcircle \\
        \hline \cite{Li23} & \solidcircle & \emptycircle & \solidcircle & \emptycircle \\
        \hline \cite{Mueller22}* & \solidcircle & \emptycircle & \solidcircle & \solidcircle \\
        \hline \cite{Zhu22} & \solidcircle & \emptycircle & \emptycircle & \solidcircle \\
        \hline \cite{Tao22} & \emptycircle & \solidcircle & \solidcircle & \emptycircle \\
        \hline \cite{Faez21} & \solidcircle & \emptycircle & \emptycircle & \solidcircle \\
        \hline \cite{Xia21} & \emptycircle & \halfcircle & \solidcircle & \solidcircle \\
        \hline \cite{Bonifati20} & \solidcircle & \emptycircle & \emptycircle & \solidcircle \\
        \hline \cite{Casas-Roma17s} & \halfcircle & \emptycircle & \solidcircle & \solidcircle \\
        \hline \cite{Ji17} & \halfcircle & \solidcircle & \halfcircle & \solidcircle \\
        \hline \cite{Hay16} & \emptycircle & \solidcircle & \solidcircle & \emptycircle \\
        \hline \cite{Ji15} & \emptycircle & \halfcircle & \halfcircle & \solidcircle \\
        \hline
    \end{tabular}
    \caption{Related work to the field of DP graph release, presented in reverse chronological order. Key: * = Preprint, \\ {\small \solidcircle} = Coverage, {\small \halfcircle} = Limited coverage, {\small \emptycircle} = No coverage.}
    \label{tab:related-work}
\end{table}

Liu et al.~(2024)~\cite{Liu24} introduced the \textit{Private Graph Benchmark} (PGB), an open-source framework for evaluating graph privatisation mechanisms, aiming to unify approaches across diverse privacy definitions. Their empirical evaluation covers methods such as DP-dK~\cite{Wang13}, TmF~\cite{Nguyen15}, PrivSKG~\cite{Mir12}, PrivHRG~\cite{Xiao14}, PrivGraph~\cite{Yuan23}, and DGG~\cite{Qin17}. However, most of these are either baseline methods or over a decade old, and deep learning techniques are deliberately excluded. This raises concerns about the work's currency and relevance to modern applications.

From a theoretical perspective, Li et al.~\cite{Li23} provide a comprehensive survey categorising existing methods by intended use (query answering vs. graph dataset publication) and by privacy guarantees (provable vs. non-provable). Mueller et al.~\cite{Mueller22} extend this to privacy challenges in GNNs and related learning tasks for graph-structured data. Earlier works by Casas-Roma et al.~\cite{Casas-Roma17s} and Ji et al.~\cite{Ji17}, surveyed graph-modification techniques for general anonymisation, without necessitating formal privacy guarantees. While these surveys consolidate methodological perspectives, their discussions of utility remain limited and fragmented, with inconsistent evaluation metrics that impede sound comparison. Consequently, practitioners face challenges in assessing the real-world applicability, validity, and utility of proposed approaches.

In practical terms, evaluating utility in graph privacy remains challenging~\cite{Ji17}, motivating systems such as \textit{SecGraph}~\cite{Ji15} and \textit{DPGraph}~\cite{Xia21}. While both offer sound frameworks, they focus on query-answering algorithms rather than graph release, and \textit{SecGraph} broadly addresses anonymisation without %
formal privacy guarantees, limiting applicability to modern privacy requirements.

Our work also aligns with benchmarking initiatives, such as \textit{DPBench}~\cite{Hay16}, which introduced principles for sound end-to-end private evaluations. Using these principles, Tao et al.~\cite{Tao22} benchmarked and evaluated DP synthetic data generation algorithms. More recent contributions include a theoretical survey by Hu et al.~\cite{Hu24}, and a benchmarking study by Chen et al.~\cite{Chen25}. However, these efforts target tabular data rather than graph-structured data.

Graph synthesis has a long history in network science~\cite{Bonifati20} and recent advances in generative deep learning have extended this to graphs, as discussed in surveys by Faez et al.~\cite{Faez21}, Guo et al.~\cite{Guo23}, and Zhu et al.~\cite{Zhu22}; however, privacy concerns such as the memorisation of training data remain largely overlooked. 

To the best of our knowledge, no prior work addressing DP graph release approaches explicitly additionally synthesises vulnerabilities and practical goals, despite these being most critical in real-world settings. Our systemisation provides further value due to its flexibility; its use depends entirely on practitioner-defined requirements, thus making it adaptable to ever-evolving contexts beyond the example demonstrations in this work. 

Our evaluation shares some similarity with PGB~\cite{Liu24} but differs in key ways: (i) inclusion of deep learning techniques; (ii) emphasis on recent mechanisms, rather than decade-old baselines, with only two overlapping methods; and (iii) broader utility metrics and downstream tasks for richer assessment. While our framework is general, evaluations are limited to social networks and are intended as illustrative instantiations rather than comprehensive.

\paragraph{Background on Differential Privacy}\label{background.differential-privacy}
Differential Privacy~\cite{Dwork06} (DP) is a mathematical framework for providing rigorous, quantifiable privacy guarantees for datasets and computations performed on them. DP has become a foundational concept in data privacy, widely adopted in both industry and research~\cite{Erlingsson14, Apple17, Abowd18}.

A randomised algorithm $A : \mathcal{X}^n \mapsto \mathcal{Y}$ is $\epsilon$-\textit{differentially private}, if for all neighbouring datasets $X, X' \in \mathcal{X}$ and $Y \subseteq \mathcal{Y}$,
\begin{equation}\label{eq:differential-privacy}
    \text{Pr}[A(X) \in Y] \le e^\epsilon \text{Pr}[A(X') \in Y],
\end{equation}
where $X$ and $X'$ differ in exactly one record, and $\epsilon$ is the privacy budget. Intuitively, noise is added to the output of $A$, which operates on a sensitive dataset $X$ to inhibit inferences about any singular record. The noise depends on two parameters: the privacy budget $\epsilon$, and the sensitivity $\Delta A$. The privacy budget quantifies the allowed ``privacy loss''; a smaller $\epsilon$ means stronger privacy guarantees (but lower accuracy), and vice versa for a larger $\epsilon$. The \textit{Global $\ell_p$ Sensitivity}, $\Delta A = \max_{X, X'} || A(X) - A(X')||_p$, expresses the largest possible distance between neighbouring datasets. The choice of $p$-norm depends on the specific noise mechanism used.

Initially designed for tabular data, DP has been adapted to accommodate graphs. Here, the \textit{neighbouring dataset} notion is adapted to various privacy targets, including edges (edge-DP~\cite{Karwa11}), nodes (node-DP~\cite{Kasiviswanathan13a}), connected components (partition-DP~\cite{Task14}), and attributes ((node/edge)-attribute-DP~\cite{Blocki13}). 
As they overwhelmingly dominate the literature, we focus on edge- and node-DP, for which we provide formal definitions in App.~\ref{formal-definitions}.

\section{Systemisation}\label{systemisation}
\paragraph{Scope and Methodology} Our systemisation focuses on methods for releasing entire graph datasets privately (not just statistics, properties, or latent features), specifically using indistinguishability-based privacy notions such as DP and its extensions. Other privacy notions, such as anonymity (e.g. $k$-anonymity~\cite{Samarati98}, $l$-diversity~\cite{Machanavajjhala07}), information-theoretic definitions~\cite{Bloch21} or multiparty computation~\cite{Lindell20} are outside our scope. Similarly, private GNN and Federated Learning models are only included if they are specifically designed to output private versions of graph datasets. 

To collate existing works, we conducted an exhaustive search using the following keywords and their variants: \{blowfish, differential, dynamic, edge, generation, graph, link, local, network, node, privacy, publication, pufferfish, relation, release, synthesis, temporal, vertex\}. These were derived from common terms associated with graphs, data publishing, differential privacy, and established variants. We include works published within the last decade (2015--2025), supplemented by earlier studies deemed highly influential or foundational to the field. Our systemisation is then applied to this collection, with the results summarised in Table~\ref{tab:sok}.

\begin{figure*}[t]
  \centering
  \includegraphics[width=\textwidth]{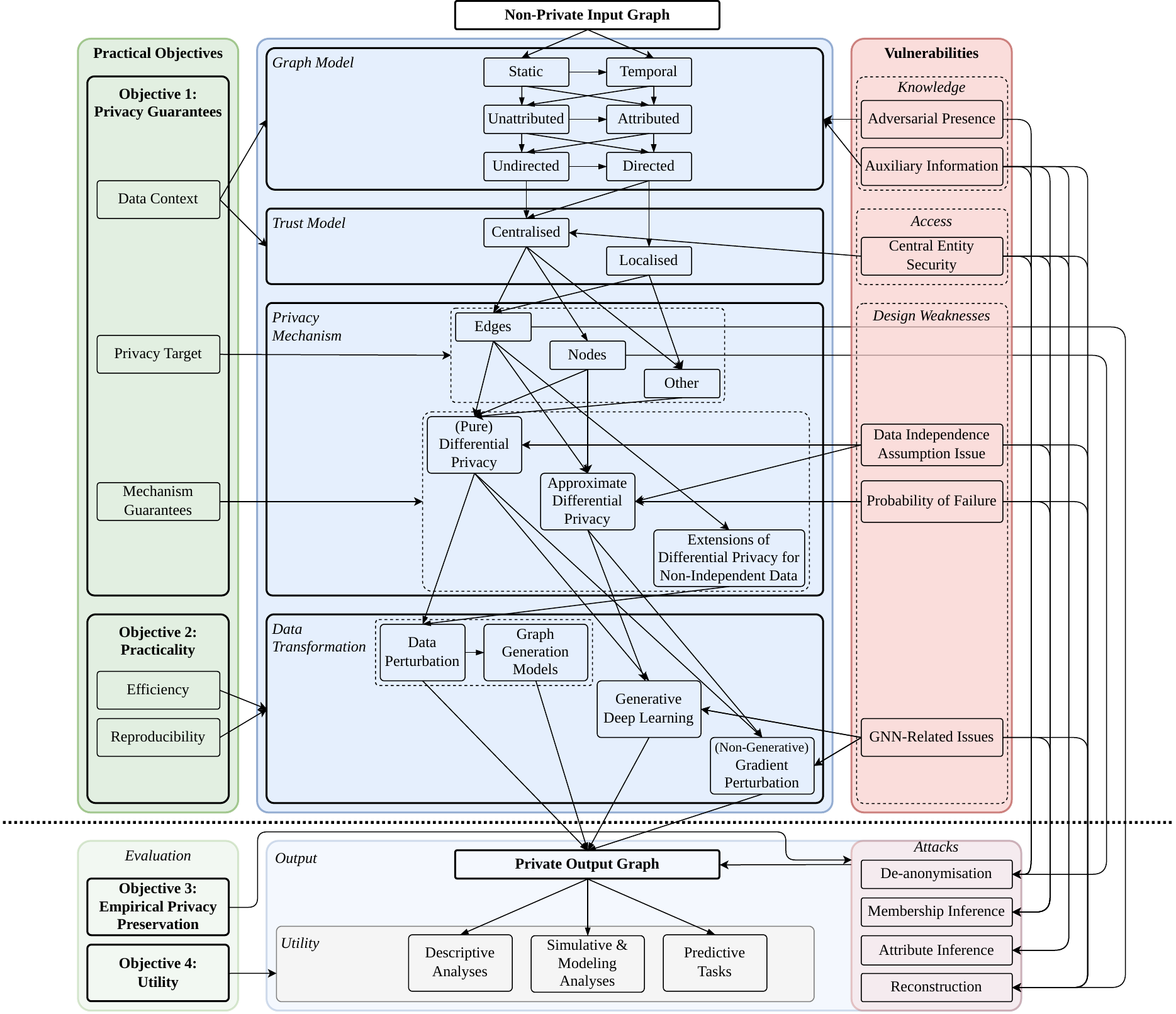}
  \caption{Systemisation of DP graph release methods (central, blue), with two practical aspects: vulnerabilities (right, red), and practitioner objectives (left, green). The reader is encouraged to parse this as a flowchart from top to bottom, along the central blue column from sensitive input to private output.}
  \label{systemisation-diagram}
\end{figure*}

\paragraph{Structure} Central to our systemisation is an end-to-end pipeline (shown in the central blue column of Fig.~\ref{systemisation-diagram}) that spans from non-private input to private output, used to categorise existing approaches. The pipeline consists of four layers:
\begin{enumerate}[label=L\arabic*:, leftmargin=*]
    \item The \textit{graph model}, which identifies the applicable graph types for the approach;
    \item The \textit{trust model}, which specifies the assumed trust relationship between the individuals and data publishers;
    \item The \textit{privacy mechanism}, comprising the \textit{privacy target}, (elements over which privacy should be applied, and the \textit{privacy definition} (formalises privacy guarantees); and
    \item The \textit{privacy-enforcing transformation}, which describes the computational process applied to produce the output graph.
\end{enumerate}
Each method we reviewer represents a path through these layers, from top to bottom. This modular view enables a targeted examination of each component of any given graph privatisation method, revealing how design choices influence utility, where consistent strengths and weaknesses emerge, and where future efforts may be directed.

We also integrate a vulnerability dimension (right red column in Fig.~\ref{systemisation-diagram}), categorising risks as knowledge-based, access-based, or design-related, any of which may lead to attacks on the private output. To the best of our knowledge, no previous systemisation explicitly integrates this dimension; however, we find it critical for understanding failure points under different threat models.

To complement this, we introduce a practitioner-oriented, object-ive-based framework (left green column in Fig.~\ref{systemisation-diagram}) mapped to the pipeline layers. It helps practitioners, even those without domain expertise, select and apply DP-based graph release mechanisms by clarifying guarantees and limitations. Such translation is essential given the well-documented usability challenges of DP~\cite{Ivoline24} and the growing number of mechanisms, which make it difficult for practitioners to identify those most aligned with their goals.

Categories below the dotted line in Fig.~\ref{systemisation-diagram} (\textit{O3} and \textit{O4}) require evaluation for a sound comparison. Practitioners operate under diverse conditions (e.g., data domain, privacy/utility/computational needs). In the literature, evaluation practices vary widely, with inconsistent methodologies, privacy budgets, and metrics, making direct comparisons difficult. This lack of standardisation hinders data practitioners' ability to assess utility and robustness against attacks. Our approach advocates evaluation tailored to practitioner-defined contexts, producing results aligned with their privacy and utility goals. These evaluation-related components are discussed in Secs.~\ref{systemisation.objectives.empirical-privacy}--\ref{systemisation.objectives.utility}; two illustrative scenarios and empirical demonstrations are presented in 
Secs.~\ref{evaluations.scenario-1} and \ref{evaluations.scenario-2} respectively.

\begin{table*}
    \small
    \centering
    \begin{tabular}{c|ccccc}
        \hline \textbf{Work} & \textbf{Graph Model} & \textbf{Trust Model} & \textbf{Privacy Mechanism} & \textbf{Transformation} \\
        \hline HMG \cite{Li25} & Undirected, Temporal, Unattributed & Localised & Edge-DP & Perturbation then Generation \\
        \hline PrivDPR \cite{Zhang25} & Directed, Static, Unattributed & Centralised & Node-ADP & Perturbation of Gradients \\
        \hline Zou et al. \cite{Zou25} & Undirected, Static, Unattributed & Centralised & Edge-ADP & Perturbation \\
        \hline DPCAG \cite{Zhang24} & Directed, Static, Unattributed & Centralised & Edge-DP & Perturbation then Generation \\
        \hline (``Global'') Brito et al. \cite{Brito23} & Undirected, Static, Weighted & Centralised & (Edge \& Weight)-DP & Perturbation then Generation \\
        \hline (``Local'') Brito et al. \cite{Brito23} & Undirected, Static, Weighted & Localised & (Edge \& Weight)-DP & Perturbation then Generation \\
        \hline PPDU \cite{Hou23} & Undirected, Temporal, Unattributed & Localised & Edge-DP & Perturbation \\
        \hline $\pi_v$, Jian et al. \cite{Jian23} & Undirected, Static, Unattributed & Centralised & Node-DP & Perturbation \\
        \hline $\pi_e$, Jian et al. \cite{Jian23} & Undirected, Static, Unattributed & Centralised & Node-ADP & Perturbation \\
        \hline CGT \cite{Yoon23} & Directed, Static, Attributed & Centralised & Edge-ADP & Generative Deep Learning \\
        \hline PrivGraph \cite{Yuan23} & Undirected, Static, Unattributed & Centralised & Edge-DP & Perturbation then Generation \\
        \hline Chicha et al. \cite{Chicha21a} & Undirected, Temporal, Unattributed & Centralised & Edge Blowfish Privacy & Perturbation \\
        \hline PSG \cite{Huang21} & Undirected, Static, Unattributed & Localised & Edge-DP & Perturbation then Generation \\
        \hline WGPA \cite{Ning21} & Undirected, Static, Weighted & Centralised & (Edge \& Weight)-DP & Perturbation then Generation \\
        \hline DPGVAE \cite{Yang21} & Directed, Static, Unattributed & Centralised & Edge-ADP & Generative Deep Learning \\
        \hline DPGGAN \cite{Yang21} & Directed, Static, Unattributed & Centralised & Edge-ADP & Generative Deep Learning \\
        \hline CAGMDP \cite{Chen20} & Undirected, Static, Attributed & Centralised & (Edge \& Node-Att.)-DP & Perturbation then Generation \\
        \hline Eli\'{a}\v{s} et al. \cite{Elias20} & Undirected, Static, Unattributed & Centralised & Edge-ADP & Perturbation of Gradients \\
        \hline MPDC-dK \cite{Iftikhar20} & Undirected, Static, Unattributed & Centralised & Edge-DP & Perturbation \\
        \hline VB-WNDP \cite{Wang20} & Directed, Static, Weighted & Centralised & (Edge 
        \& Weight)-DP & Perturbation then Generation \\
        \hline AsgLDP \cite{Wei20} & Undirected, Static, Attributed & Localised & Edge-DP & Perturbation then Generation \\
        \hline Yang et al. \cite{Yang20} & Undirected, Static, Unattributed & Localised & Edge-DP & Perturbation then Generation \\ 
        \hline PrivCom \cite {Zhang20} & Undirected, Static, Unattributed & Centralised & Node-ADP & Perturbation of Gradients \\
        \hline PHDP \cite{Gao19} & Undirected, Static, Unattributed & Centralised & Edge-DP & Perturbation then Generation \\
        \hline LMBCI \cite{Wang19} & Undirected, Static, Weighted & Centralised & Weight-DP & Perturbation then Generation \\
        \hline DP-FT \cite{Zhu19} & Undirected, Static, Unattributed & Centralised & Degree-DP & Perturbation then Generation \\
        \hline Gao et al. \cite{Gao18} & Undirected, Static, Unattributed & Localised & (Grouped Node)-DP & Perturbation then Generation \\
        \hline MB-CI \cite{Li17} & Undirected, Static, Weighted & Centralised & Weight-DP & Perturbation then Generation \\
        \hline LDPGen \cite{Qin17} & Directed, Static, Unattributed & Localised & Edge-DP & Perturbation then Generation \\
        \hline TriCycLe \cite{Jorgensen16} & Undirected, Static, Attributed & Centralised & (Edge \& Node-Att.)-DP & Perturbation then Generation \\
        \hline TmF \cite{Nguyen15} & Undirected, Static, Unattributed & Centralised & Edge-DP & Perturbation then Generation \\
        \hline DER \cite{Chen14} & Undirected, Static, Unattributed & Centralised & Dep. Edge-DP & Perturbation then Generation \\
        \hline PrivHRG \cite{Xiao14} & Undirected, Static, Unattributed & Centralised & Edge-DP & Perturbation then Generation \\
        \hline DP-2K \cite{Wang13} & Undirected, Static, Unattributed & Centralised & Edge-ADP & Perturbation then Generation \\
        \hline Blocki et al. \cite{Blocki12} & Directed, Static, Unattributed & Centralised & Edge-DP & Perturbation \\
        \hline PrivSKG \cite{Mir12} & Undirected, Static, Unattributed & Centralised & Edge-ADP & Perturbation then Generation \\
        \hline Pygmalion \cite{Sala11} & Undirected, Static, Unattributed & Centralised & Edge-DP & Perturbation then Generation \\
        \hline
    \end{tabular}
    \caption{Considered works on DP graph release, categorised according to our proposed systemisation, presented in reverse chronological order. Abbreviations: Dep. = Dependency-aware, Att. = Attribute, (A)DP = (Approximate) Differential Privacy.}
    \label{tab:sok}
\end{table*}

\subsection{Approaches}\label{systemisation.approaches}
\subsubsection{L1: Graph Models}\label{systemisation.approaches.graph-models}
A graph model defines the type(s) of graphs that an approach is designed to handle, such as undirected or directed, static or temporal, and semantically attributed or unattributed. Varying real-world applications necessitate different graph models. For example, in social networks, undirected edges represent mutual relationships (e.g. friendships), while directed edges capture asymmetric interactions (e.g. a user following another), leading to distinct dynamics and properties.

Adding attributes or temporal information enhances the representational capacity of graphs. Attributes encode contextual details, such as demographic information or community affiliations, while temporal dimensions capture dynamic processes, such as relationship evolution or influence propagation. While these elements can enhance realism, they heighten privacy risks by exposing sensitive details absent in unattributed or static graphs.

Graph models characterise the generalisability of approaches, as they follow a hierarchy: directed graphs generalise undirected graphs, temporal generalises static (the case of a single time-step), and attributed generalises unattributed (by adding trivial attributes). Methods capable of handling directed, temporal, and attributed graphs are thus applicable across diverse scenarios.

From Table~\ref{tab:sok}, we observe a clear progression. Earlier works focus on simple graphs (undirected, unattributed, static), often grounded in theoretical foundations. Recent studies increasingly address attributed and temporal graphs, reflecting the growing scale, richness, and complexity of contemporary graph data.

\subsubsection{L2: Trust Models}\label{systemisation.approaches.trust-models}
We assume individuals and their data are the subjects of protection, though this generalises to any entity whose data requires safeguarding. \textit{Individuals} generate data, which a \textit{data publisher} collects to release as a dataset for a \textit{data recipient}. Existing graph privacy literature consists of two well-established trust models: centralised and localised. In both, individuals do not directly trust the recipient, so the published dataset must protect privacy while ensuring the recipient trusts the publisher’s accuracy.

\paragraph{Centralised} Individuals trust the publisher and share their raw data. The publisher applies a privacy mechanism before releasing a protected dataset. This model represents the de facto standard for publishing privacy-preserving data.

\paragraph{Localised} Individuals do not trust the publisher and apply privacy mechanisms locally before sharing data. The publisher aggregates and optionally post-processes this privatised data. This model offers stronger privacy but often reduces utility and introduces practical constraints, such as communication inefficiencies.

The original DP definition assumes the centralised model, making it the de facto standard. A key variant, local-DP~\cite{Kasiviswanathan11w}, adapts DP to the localised model. Here, the lack of trust in individuals towards a central curator implies a stronger notion of privacy than in the centralised setting. However, this stronger guarantee often comes at the cost of reduced utility.
For example, Kasiviswanathan et al.~\cite{Kasiviswanathan11w} proved that some local-DP learning algorithms require exponentially more data and are strictly less powerful than their centralised counterparts. The localised model induces a client-server architecture, where each client corresponds to a graph node, introducing practical constraints, e.g., communication inefficiencies. The increased difficulty of mitigating the localised model's adverse effects on utility and efficiency explains why, as shown in Table~\ref{tab:sok}, far fewer works target this trust model, leaving it an open area for future research. Other trust models, such as the shuffle model~\cite{Cheu19} (which may see application in attributed graphs where nodes could have their feature vectors shuffled), remain largely unexplored and present opportunities for further investigation.

\subsubsection{L3: Privacy Mechanism}\label{systemisation.approaches.privacy-mechanism}
This considers the privacy target, i.e., the graph elements to protect (e.g., nodes, edges, attributes), and the privacy definition that formalises this. We distinguish between these definitions to account for differing guarantees they provide.

\paragraph{(Pure) Differential Privacy}
For graphs, DP is typically instantiated as edge-DP~\cite{Blocki12} or node-DP~\cite{Kasiviswanathan13a}, protecting edges or nodes, respectively. Most work focusses on edge-DP, while node-DP remains largely unaddressed. This imbalance reflects the relative ease of preserving utility in edge-DP.  Sensitivity (Sec.~\ref{background.differential-privacy}) quantifies the maximum distance between neighbouring datasets. In edge-DP, neighbouring graphs differ by the addition or removal of an edge, inducing an $O(1)$ sensitivity for many typical queries. In node-DP, they differ by the addition or removal of a node and all incident edges. In the worst case, a node is connected to all others, yielding an $O(|V|)$ sensitivity for several typical queries~\cite{Kasiviswanathan13a}. Achieving node-DP thus requires far more noise than for edges with the same privacy budget, degrading accuracy and utility.

Another concern is that, for DP guarantees (Eq.~\ref{eq:differential-privacy},~\cite{Dwork06}) to hold, records are assumed to be independent. In edge-DP, edges are supposed to exist independently, and likewise nodes in node-DP. However, since graphs are inherently relational, this assumption rarely holds in practice. Kifer and Machanavajjhala~\cite{Kifer11} illustrated this through social network simulations, showing that the dependent structure of graphs undermines this assumption. In their example, if two previously isolated communities become connected through a new link, additional links are more likely to form across the communities; however, these new links are dependent on the existence of the first. They argue that privacy should limit an attacker’s ability to infer participation, regardless of explicit presence in the data. This dependency challenge remains unresolved in graph privacy.

\paragraph{Approximate Differential Privacy}
Approximate DP (ADP) is a relaxed variant of pure DP that introduces a parameter $\delta$, which quantifies the probability of failure, i.e. of the claimed guarantees (as per pure DP) not holding. With the same definitions as Eq.~\ref{eq:differential-privacy}, 
\begin{equation}\label{eq:approximate-differential-privacy}
    \text{Pr}[A(X) \in Y] \le e^\epsilon \text{Pr}[A(X') \in Y] + \delta.
\end{equation}
Typically, $\delta$ is set to very small values such as $10^{-5}$~\cite{Zhang20}. Typically, the Gaussian mechanism~\cite{Dwork14} is used here, which incurs less noise for higher-dimensional outputs, and has seen widespread adoption in deep learning contexts upon the development of DP-SGD~\cite{Abadi16}. Consequently, all methods in Table~\ref{tab:sok} that make use of deep learning and other optimisation methods provide ADP guarantees. However, the same concerns regarding node sensitivity and the (non-)independence issue remain with this definition.

\paragraph{Extensions of Differential Privacy for Non-Independent Data}
To address the (non-)independence issue noted, several extensions to standard DP have been proposed. The most na\"ive is \textit{Group Privacy}~\cite{Dwork06}, a corollary of DP that can be used to prevent the leakage of sensitive information from a collection of records rather than singularly. If a mechanism $A$ satisfies $\epsilon$-DP, then for any group of $k$ records, $A$ satisfies $(k\epsilon)$-DP. This allows assuming correlations within groups of bounded size and scaling the privacy budget based on a `correlation parameter'. This approach was used by Chen et al. in DER~\cite{Chen14}, in which they consider correlation parameters $\le 25$.

Other variants of DP include Dependent DP~\cite{Liu16}, an alternative formulation with the same name by Zhao et al.~\cite{Zhao17}, and Bayesian DP by Yang et al.~\cite{Yang15}. Among these, the framework that has garnered the most attention is Pufferfish Privacy (PP), a generalisation of DP proposed by Kifer and Machanavajjhala~\cite{Kifer14}.

PP requires the following definitions: a set of potential secrets $\mathcal{S}$ is the privacy target, a set of discriminative pairs $S_{\text{pairs}} \subseteq \mathcal{S} \times \mathcal{S}$ explicitly specifies which pairs of secrets should be indistinguishable from each other, a set of data evolution scenarios $\mathcal{D}$ captures assumptions on how the data was generated (thus allowing to account for dependencies and correlations), and a privacy budget $\epsilon$. An algorithm $A$ satisfies $\epsilon$\textit{-Pufferfish($\mathcal{S}, S_\text{pairs}, \mathcal{D}$) Privacy} if for all possible outputs $y \in \text{Range}(A)$, for all $(s_i, s_j) \in S_\text{pairs}$, and for all distributions $\theta \in \mathcal{D}$ where $\text{Pr}[s_i | \theta] \neq 0$ and $\text{Pr}[s_j | \theta] \neq 0$, we have:
\begin{equation}
\begin{aligned}
\text{Pr}[A(X) = y | s_i, \theta] \le e^{\epsilon} \text{Pr}[A(X) = y | s_j, \theta], \\
\text{Pr}[A(X) = y | s_j, \theta] \le e^{\epsilon} \text{Pr}[A(X) = y | s_i, \theta],
\end{aligned}   
\end{equation}
where $X$ is a dataset generated from $\theta$. While PP assumes a domain expert can specify secrets and probability distributions that define the data evolution scenarios, subsequent developments have been largely theoretical. Song et al.~\cite{Song17} proposed the Wasserstein and Markov Quilt mechanisms for achieving PP, proving that these incur less noise than group privacy. However, their practicality is hindered by the intractability of the Bayesian semantics underlying PP~\cite{Yang15, Liu22}. Recently, Pierquin et al.~\cite{Pierquin24} introduced Rényi-PP, which reinstates the validity of post-processing (absent in general PP) and extends its applicability to iterative settings, such as private convex optimisation. To date, PP has been applied to graphs only by Shafieinejad et al.~\cite{Shafieinejad22}, who studied private histogram queries and set unions on communication graphs.

Blowfish Privacy (BP)~\cite{He14}, extends PP into a policy-based framework, offering more intuitive control over privatisation. A policy specifies both the set of secrets (as in PP) and constraints representing known properties of the data. At its core is a discriminative secret graph, where nodes represent records, and edges connect records that must remain indistinguishable from one another. Formally, a policy is defined as a triple $P = (\mathcal{T}, G, \mathcal{I}_Q)$ where $\mathcal{T}$ is the domain of records, $G = (V, E)$ where $V \subseteq \mathcal{T}$ and $E \coloneqq S_\text{pairs}$ is the secret graph, and $\mathcal{I}_Q$ is the set of possible databases under constraints $Q$. A randomised algorithm $A$ satisfies $(\epsilon, P)$-\textit{Blowfish Privacy} if, for all possible outputs $Y \subseteq \text{Range}(A)$, we have:
\begin{equation}
    \text{Pr}[A(X) \in Y] \le e^\epsilon \text{Pr}[A(X') \in Y],
\end{equation}
where $X, X' \in \mathcal{I}_Q$ are neighbouring datasets. Intuitively, two datasets are considered neighbouring if their corresponding secret graphs $ G$ and $ G'$ induce minimal graph edit distances~\cite{Sanfeliu83}, subject to $Q$. As an extension of PP, BP is also a generalisation of DP.

Applications of BP to graphs remain limited. Nassar et al.~\cite{Nassar20} utilised BP for private histogram queries in communication graphs with heterogeneous privacy guarantees. Chicha et al.~\cite{Chicha21a} proposed a mechanism for publishing temporal graphs under policies that capture temporal correlations between edges.

\subsubsection{L4: Privacy-Enforcing Transformation}\label{systemisation.approaches.transformation}
This captures the transformation applied to the data to achieve privacy, i.e., the core computational process behind the privacy mechanism. We identify four main strategies used in existing works:

\paragraph{Data Perturbation}
The most straightforward strategy is to perturb the input graph directly. For the methods in Table~\ref{tab:sok}, this typically involves rearranging or transforming the adjacency matrix, followed by any required corrections for the graph to remain valid. This strategy often yields simpler algorithms, as no additional work is needed. Many earlier works, including most $k$-anonymous approaches, as surveyed in~\cite{Casas-Roma17s}, follow this approach.

\paragraph{Perturbation then Generation}
More commonly, a set of graph statistics is perturbed and then used as input for various graph generation models. These generative models may be newly proposed, e.g.,~\cite{Chen14, Nguyen15, Gao19}, or drawn from established network science literature. Examples include the Attributed Graph Model~\cite{Pfeiffer14} (used in~\cite{Zhang24, Chen20, Wei20, Jorgensen16}), the Chung-Lu model~\cite{Aiello01} (e.g. \cite{Yuan23, Chen20, Jorgensen16}), the Block Two-Level Erdös–Rényi model~\cite{Seshadhri12} (e.g.~\cite{Qin17}), the Hierarchical Random Graph model~\cite{Clauset08} (e.g.~\cite{Gao18, Xiao14}), the Stochastic Kronecker Graph model~\cite{Leskovec10}, and the Havel-Hakimi algorithm~\cite{Hakimi62} (e.g.~\cite{Wang13, Sala11}). This approach offers greater expressivity than direct perturbation but has declined with the rise of deep learning.

\paragraph{Differentially Private Generative Deep Learning}
Differentially Private Stochastic Gradient Descent (DP-SGD)~\cite{Abadi16} incorporates DP into model training. Each gradient descent step is bounded by clipping the $\ell_2$-norm, and Gaussian noise is added to ensure that both the model parameters and any derived outputs satisfy DP. With the surge of generative deep learning for graph generation tasks~\cite{Zhu22}, recent studies have applied DP-SGD to train advanced models, such as Transformers~\cite{Vaswani17} (e.g.~\cite{Yoon23}), Variational Autoencoders~\cite{Kingma14vae} (e.g.~\cite{Yang21}), and Generative Adversarial Networks~\cite{Goodfellow14} (e.g.~\cite{Yang21}).

\paragraph{Differentially Private Gradient Perturbation}
Beyond generative models, DP-SGD and related gradient-privatising methods can be applied to a broader set of optimisation problems. For instance, Zhang et al.~\cite{Zhang25} use DP-SGD to devise a DP deep PageRank~\cite{Page99} method by scoring nodes and sampling a private synthetic graph, Zhang et al.~\cite{Zhang20} apply DP to Oja's method~\cite{Jain16} to approximate adjacency matrix eigen-decompositions, and Eli\'{a}\v{s} et al.~\cite{Elias20} formulate a convex optimisation problem based on graph cuts and devise a privacy-aware mirror descent~\cite{Bubeck15}.

\subsection{Vulnerabilities and Attacks}\label{systemisation.vulnerabilities}
Several studies examine attacks against DP-protected graph data~\cite{Ji15, Ji17, Mao18, Mao21, Piao21}. Thus, we provide only a brief discussion here, as our primary focus is on privacy defence methods. We categorise vulnerabilities loosely by their association with the different layers of our systemisation, grouping them by knowledge, access, and design weaknesses (see Fig.~\ref{systemisation-diagram}, right red column).

\subsubsection{Knowledge}
This captures an adversary's existing knowledge and assumptions of the target data. In graph privacy, two main categories emerge: adversarial presence and auxiliary information.

An adversarial presence occurs when the adversary can influence the data. For example, in online social networks, adversaries can create Sybil accounts~\cite{Yang14}, then re-identify them in the private graph. In graph de-anonymisation, adversaries with partial node mappings between sensitive and private graphs can launch potent seeded attacks to re-identify nodes, as opposed to comparatively weaker seed-free attacks~\cite{Ji15}. For further discussion, see~\cite{Ji17}.

Auxiliary information includes external data or assumptions that enhance attack effectiveness~\cite{Qian19}, such as record linkage~\cite{Fu14}. It also covers any assumptions about data distribution and statistics that can similarly improve attack success, as shown in~\cite{Ji15, Ji17}.

\subsubsection{Access}
This relates to the trust models adopted by various methods. As mentioned in Sec.~\ref{systemisation.approaches.trust-models}, the centralised model requires a higher degree of trust than the local model. Consequently, ensuring the security of the centralised entity acting as the aggregator is critical. If this is compromised, sensitive raw data could leak.

\subsubsection{Design Weaknesses}
These vulnerabilities stem from limitations in the selection of privacy mechanisms and in algorithm design. As noted in Sec.~\ref{systemisation.approaches.privacy-mechanism}, common issues include assuming independent records or allowing a non-zero probability of failed protection guarantees. Both can lead to underestimating the required noise and inadequate data masking.

Furthermore, all deep learning-based methods~\cite{Yang21, Yoon23, Zhang25} apply DP only to later feed-forward layers due to the difficulty of applying DP-SGD to convolutional settings~\cite{Abadi16}, including GNNs~\cite{Mueller22}. Isolating the influence of a single node in a graph is inherently challenging due to message passing, which complicates per-example gradient computation. Without careful design, independence assumptions are violated, as dependent information propagates through random walks~\cite{Grover16} and GCN-based embeddings~\cite{Kipf17}.

\subsubsection{Attacks}\label{systemisation.vulnerabilities.attacks}
We also present a succinct categorisation of existing attacks in 
Table~\ref{tab:attacks},
grouped by attack type and privacy target, to help identify which threat models should be considered. In summary, de-anonymisation attacks primarily concern nodes, as these are mainly used to model people and their identities. Reconstruction attacks attempt to recover the graph structure, thereby additionally recovering edge information. Attribute inferences can apply either to node or edge attributes, and similarly, membership inference can be conducted over nodes or edges.

\begin{table}
    \small
    \centering
    \begin{tabular}{c|cc}
         \hline \textbf{Work} & \textbf{Attack Type} & \textbf{Privacy Target} \\
         \hline \cite{Azogagh25} & Reconstruction & Nodes \& Edges \\
         \hline \cite{Lu24} & De-anonymisation & Nodes \\
         \hline \cite{Yuan24} & Attribute Inference & Node Attributes \\
         \hline \cite{Desai21} & De-anonymisation & Nodes \\
         \hline \cite{Zhang21} & De-anonymisation & Nodes \\
         \hline \cite{Li20} & De-anonymisation &  Nodes \\
         \hline \cite{Zhang20d} & De-anonymisation & Nodes \\
         \hline \cite{Zhang20t} & (Edge) Membership Inference & Edges \\
         \hline \cite{Mao19} & Attribute Inference & Node Attributes \\
         \hline \cite{Gao19d} & De-anonymisation & Nodes (Temporal Graph) \\
         \hline \cite{Sun19} & De-anonymisation & Nodes \\
         \hline \cite{Zhang19} & De-anonymisation & Nodes (Attributed Graph) \\
         \hline \cite{Hu19} & De-anonymisation & Nodes \\
         \hline \cite{Wu18} & De-anonymisation & Nodes \\
         \hline \cite{Jia17} & Attribute Inference & Node Attributes \\
         \hline \cite{Ji17d} & De-anonymisation & Nodes \\
         \hline \cite{Qian17} & De-anonymisation & Nodes \\
         \hline \cite{Lee17} & De-anonymisation & Nodes \\
         \hline \cite{Wong16} & Attribute Inference & Node Attributes \\
         \hline \cite{Ji16} & De-anonymisation & Nodes \\
         \hline \cite{Gong16} & Attribute Inference & Node Attributes \\
         \hline \cite{Erdos14} & Reconstruction & Nodes \& Edges \\
         \hline \cite{Sharad14} & De-anonymisation & Nodes \\
         \hline \cite{Korula14} & De-anonymisation & Nodes \\
         \hline \cite{Ji14} & De-anonymisation & Nodes \\
         \hline \cite{Nilizadeh14} & De-anonymisation & Nodes \\
         \hline \cite{Yartseva13} & De-anonymisation & Nodes \\
         \hline \cite{Pedarsani13} & De-anonymisation & Nodes \\
         \hline \cite{Fire13} & (Edge) Membership Inference & Edges \\
         \hline \cite{Srivatsa12} & De-anonymisation & Nodes \\
         \hline \cite{Mislove10} & Attribute Inference & Node Attributes \\
         \hline \cite{Wu10} & Reconstruction & Nodes \& Edges \\
         \hline \cite{Lindamood09} & Attribute Inference & Node Attributes \\
         \hline \cite{Narayanan09} & De-anonymisation & Nodes \\
         \hline \cite{Clauset08} & (Edge) Reconstruction & Edges \\
         \hline \cite{Comellas08} & Reconstruction & Nodes \& Edges \\
         \hline \cite{Korolova08} & (Edge) Membership Inference & Edges \\
         \hline \cite{Bhagat07} & Attribute Inference & Node Attributes \\
         \hline \cite{Backstrom07} & De-anonymisation & Nodes \\
         \hline
    \end{tabular}
    \caption{Considered works on attacks against privately released graph data, presented in reverse chronological order.}
    \label{tab:attacks}
\end{table}

\subsection{Practitioner Objectives}\label{systemisation.objectives}
The final component is a practitioner-oriented, objective-based framework for curating suitable approaches, adaptable to specific applications and requirements. It consists of four key objectives: the privacy guarantees they provide (given context), their practicality, their empirical protection against known attacks, and the utility they provide (see Fig.~\ref{systemisation-diagram}, left green column). This assists practitioners in answering the question, ``Which private graph release mechanisms are feasible for us to apply, given our requirements?'' This motivation arises from the growing number of available mechanisms, which makes it difficult for practitioners to identify those aligned with their goals. It also clarifies the nature and validity of the guarantees promised by each mechanism, mitigating misleading claims of data protection. We link each objective to the relevant layer(s) of our central pipeline, clarifying relationships between practical goals, and mechanism components and their vulnerabilities. \textit{O1} and \textit{O2} can be employed pre-evaluation to filter and select candidate mechanisms. Then, evaluations are conducted to answer \textit{O3} and \textit{O4} in accordance with the practitioner's specifications.

\subsubsection{O1: Privacy Guarantees}\label{systemisation.objectives.privacy-gurantees}
This objective broadly addresses the privacy guarantees offered by a mechanism and the conditions under which they hold. We divide this objective into two sub-objectives: \textit{context}, and \textit{mechanism target and properties}.

\paragraph{Context} This refers to the specific graph model(s) that the mechanism is designed to address, as well as the trust assumptions that underpin its design. As mentioned in Sec.~\ref{systemisation.approaches}, these concepts are essential in determining the applicability of a mechanism. For example, mechanisms that only handle undirected graphs cannot be applied to directed graphs. Similar distinctions arise between temporal and static graphs, and between unattributed and attributed graphs. Equally important is the trust model; specific architectures may require localised mechanisms (e.g., local DP), whereas others permit centralised approaches. Also importantly, some mechanisms are tailored to specific domains (e.g. social networks~\cite{Jorgensen16, Qin17, Chen20}), and may not perform well in other contexts. This sub-objective corresponds to layers \textit{L1} and \textit{L2} in our systemisation of approaches.

\paragraph{Mechanism Target and Properties} This sub-objective relates to the privacy target and its properties. Practitioners must clearly define the privacy target (e.g., edges, nodes, attributes) and select mechanisms to protect it, while understanding the guarantees offered. For example, DP offers strong, provable guarantees, but assumes independent record generation~\cite{Dwork06, Kifer11}. Several mechanisms use ADP rather than (pure) DP~\cite{Zhang25, Yang21, Zhang20}; while this definition can improve utility, the trade-off, which increases the risk of failure (quantified by $\delta$), should be clearly acknowledged. Misinterpreting these conditions can lead to privacy breaches or misleading claims. This sub-objective aligns with \textit{L3}.

\subsubsection{O2: Practicality}\label{systemisation.objectives.practicality}
For a privacy mechanism to deliver utility, it must be readily applicable in real-world settings, rather than remain theoretical. Mechanisms that cannot be executed or implemented given the available resources ultimately provide no practical value. Accordingly, practitioners must assess which mechanisms are feasible given resource constraints. We frame this objective as comprising two sub-objectives: \textit{efficiency} and \textit{reproducibility}.

\paragraph{Efficiency} Mechanisms need to run efficiently on datasets of a size relevant to the practitioner, without exceeding available computing resources or time budgets. Otherwise, methods that appear theoretically suitable may become impractical due to excessive computation or execution times. This aspect is most influenced by \textit{L4}, which encompasses the core computational processes. Examples include training deep learning models~\cite{Yang21, Zhang25}, switching edges~\cite{Jian23, Casas-Roma17}, or executing complex graph generation procedures~\cite{Qin17, Chen20}, all of which impact efficiency.

\paragraph{Reproducibility} A mechanism must be reproducible and accessible to practitioners; otherwise, correct implementation becomes difficult, undermining both efficiency and the reliability of its claimed privacy guarantees. To meet this objective, source code should be publicly available, or detailed implementation guidance must be provided to enable accurate and confident reimplementation.

\subsubsection{O3: Empirical Privacy Preservation}\label{systemisation.objectives.empirical-privacy}
While \textit{O1} ensures privacy guarantees hold theoretically, practitioners must also validate that they protect against threats in practice. This is critical when strong assumptions are made about the data when selecting mechanisms based on \textit{O1} and \textit{O2}. After developing a threat model, a set of relevant attacks should be chosen, in accordance with the privacy target specified in \textit{O1}, to validate whether privacy is preserved in practice. To ensure reliability, the same metrics and implementation (source code) as the original publication should be used.

\subsubsection{O4: Utility}\label{systemisation.objectives.utility}
Graphs are typically published to support downstream analytical tasks, making it essential that mechanisms preserve sufficient utility to justify their release. Various metrics have been employed in the literature to quantify how well graph properties and performance on downstream tasks are preserved (e.g., modularity for community structure~\cite{Qin17} and link prediction~\cite{Yang21}); however, these metrics exhibit significant variations across studies. Thus, comparability is limited. When practitioners know in advance which downstream tasks will be performed on a published graph, they can select metrics aligned with those tasks. In the absence of such task-specific knowledge, a more comprehensive evaluation of utility is needed to guide the curation of suitable approaches. 

We categorise utility into three broad classes of common graph analytical tasks: descriptive, simulative, and predictive. Descriptive tasks compute graph statistics and structural metrics. Simulative tasks model graph dynamics and behaviour, such as influence maximisation, stochastic block models~\cite{Holland83}, and exponential random graph models~\cite{Frank86, Wasserman96}. Predictive tasks infer outcomes and trends not explicitly encoded in the graph, with learning tasks forming a key subset. The widespread adoption of GNNs attests to their utility and applicability across diverse domains.

Lastly, we highlight the principles-based framework proposed by Hay et al.~\cite{Hay16} to enable the sound and fair evaluation of DP mechanisms. These principles extend well into this domain and should be adhered to across the evaluative portion of this systemisation, as demonstrated with our examples in Sec.~\ref{evaluations.methodology}.

\section{Illustrative Evaluations}\label{evaluations}
We provide two exemplar evaluations that demonstrate our systemisation in practice. To do this, we assume the role of a social network data practitioner under two different scenarios, one focusing on edge-DP (Sec.~\ref{evaluations.scenario-1}), the other on node-DP (Sec.~\ref{evaluations.scenario-2}). In addition to the information in the following subsections, an annotated version of our systemisation (Fig.~\ref{systemisation-annotated-appendix}, App.~\ref{appendix.systemisation-annotated}) visually represents the two scenarios. Within each, we conduct evaluations accordingly to illustrate the evaluative portion of our framework. For both evaluations, we consider mechanisms that represent the SotA, published in ICORE~\cite{ICORE} A* or A-ranked venues from the last decade. Inclusion and exclusion criteria for methods evaluated are handled by applying \textit{O1} and \textit{O2}; their applications in Scenarios \num{1} and \num{2} are shown in Tables~\ref{tab:edge-DP-mechanisms} and~\ref{tab:node-dp-mechanisms} respectively.

Together, these evaluations provide a SotA benchmark of recent developments in DP graph publishing in the social networking domain, one of the most widespread applications in the literature. We emphasise, these results do not necessarily translate to other domains. Our deliberate focus on a single domain exemplifies practical instantiations of our systemisation. Accounting for domain-specific variations in utility metrics and relevant threat models requires an infeasible scale of evaluations, beyond our scope.

\subsection{Evaluation Methodology}\label{evaluations.methodology}
For both evaluations, we adopted the following methodology. Each mechanism was evaluated across twelve privacy budgets ($\epsilon$) chosen to reflect the exponential relationship: $\epsilon \in$ [0.5, 0.75, 1, 1.5, 2, 3, 4.5, 6.5, 9, 12, 16, 20]. This design maintains consistency between the amount of noise (privacy) and measured utility, thereby clarifying observable trends. Intuitively, the difference between $\epsilon = 1$ and $\epsilon = 2$ is more drastic than that between $\epsilon = 11$ and $\epsilon = 12$, which our chosen values reflect. This also enables us to efficiently cover a broader range of $\epsilon$, addressing what practitioners may consider to be ``strong'', ``moderate'', and ``weak'' privacy regimes in real-world policy settings. They approximately follow an exponential growth pattern; however, rounded values were selected to maintain comparability with existing studies. %
For each method, \num{10} trials were run across the \num{12} privacy budgets, yielding \num{120} graphs for measurement. Reported results correspond to trial averages. 

After generating the private graphs, utility and empirical privacy evaluations were performed. Due to the high computational demands of some attacks, evaluations were performed on the smallest dataset over a subset of privacy budgets: $\epsilon \in [1, 3, 9]$, with the average taken across \num{10} trials for each.

\subsection{Example Scenario 1: Edge Privacy}\label{evaluations.scenario-1}
For the first scenario, we assume the role of a practitioner in the social network analysis domain, concerned with ensuring privacy over the structural information (i.e., edges) of their data. We outline the scenario details according to the objectives in our systemisation:

\subsubsection{O1}
Originally sourced from the Stanford Network Analysis Project repository~\cite{snap}, we have four static, undirected, node-attributed social graphs: Facebook~\cite{Facebook}, LastFM~\cite{LastFM}, GitHub~\cite{GitHub}, and Brightkite~\cite{Brightkite}. To evaluate generalisability within our domain, the datasets vary in size and sub-domain (online social, music, development, and location-based, respectively). 

\begin{table}
    \small
    \centering
    \begin{tabular}{c|cccc}
    \hline        & \textbf{Facebook} & \textbf{LastFM} & \textbf{GitHub} & \textbf{Brightkite} \\ %
    & \cite{Facebook} & \cite{LastFM} & \cite{GitHub} & \cite{Brightkite} \\
    \hline \# Nodes  &  \num{4039} & \num{7624} &  \num{37700} & \num{58228}  \\
    \hline \# Edges  &  \num{88234} & \num{27806} & \num{289003} & \num{214078} \\
    \hline Density & \num{0.0108} & \num{0.0010} & \num{0.0004} & \num{0.0001} \\
    \hline H. Diameter & \num{3.2618} & \num{4.8742} & \num{3.0680} & \num{4.8982} \\
    \hline Assortativity & \num{0.0636} & \num{0.0170} & \num{-0.0752} & \num{0.0108} \\
    \hline Modularity & \num{0.8348} & \num{0.8143} & \num{0.4537} & \num{0.6862} \\
    \hline Infl. Max. Sp. & \num{58.7196} & \num{9.6470} & \num{29.7435} & \num{12.4432} \\
    \hline L.P. AUROC & \num{0.9666} & \num{0.7930} & \num{0.9134} & \num{0.8259} \\
    \hline \# Node Classes & No Labels & \num{5}* & \num{2} & No Labels \\
    \hline N.C. F1-Score & No Labels & \num{0.2426} & \num{0.6207} & No Labels \\
    \hline
    \end{tabular}
    \caption{Dataset properties. Key: * = After preprocessing. \\
    H. = Harmonic; Infl. Max. Sp. = Influence Maximisation Spread; L.P. = Link Prediction; N.C. = Node Classification.}
    \label{tab:datasets}
\end{table}

Several mechanisms that process node features \cite{Jorgensen16, Chen20} are designed for binary attributes. Following \cite{Chen20}, we limit the number of features to \num{50}. For the Facebook dataset, which %
contains missing data, we retain the \num{50} features with the fewest missing entries. The other datasets have categorical features, which we one-hot encode. To preserve end-to-end privacy, target variables are excluded during preprocessing. Instead, we heuristically select the \num{50} features with the highest entropy. The LastFM dataset initially has \num{18} target classes which we reduce to \num{5} via $k$-means clustering. This improves classification interpretability before evaluation, thereby enabling more stable and meaningful measurement of performance changes resulting from the application of privacy mechanisms.

We implement centralised and localised mechanisms. We accept a small probability of failure, $\delta = 10^{-5}$, arising from ADP. We also accept the potential weakening of guarantees from applying DP over possibly non-independent edges. Thus, empirical validation is required to assess practical privacy preservation (\textit{O3}).

\subsubsection{O2}
Mechanisms must be computationally feasible on our system: a machine with Ubuntu 24.04 LTS, an AMD EPYC 7763 64-Core processor, 500GB of RAM, and two NVIDIA RTX A6000 GPUs with 48GB of VRAM each. As per Sec.~\ref{systemisation.objectives.practicality}, they must also be reproducible. %
The first column of Table~\ref{tab:edge-DP-mechanisms} shows mechanisms available after applying \textit{O1}, and the next two show \textit{O2}. The authors of CGT~\cite{Yoon23} note an inability to achieve convergence using DP-SGD on a transformer, thus rendering the method a theoretical result and not reproducible. Eli\'{a}\v{s} et al. noted their method is implementable with a time complexity of $O(n^7)$, inefficient given the scale of our datasets. We then include satisfactorily efficient and reproducible works in our evaluation; we were unable to obtain source code for those remaining works indicated as `not reproducible'. 

\begin{table}
    \small
    \centering
    \begin{tabular}{c|ccc}
        \hline \textbf{Work} & \textbf{Privacy Mechanism} & \textbf{Efficient} & \textbf{Reproducible} \\
        \hline DPCAG~\cite{Zhang24} & Edge-DP & \solidcircle & \emptycircle \\
        \hline CGT \cite{Yoon23} & Edge-ADP & \halfcircle & \emptycircle \\
        \hline PrivGraph \cite{Yuan23} & Edge-DP & \solidcircle & \solidcircle\\
        \hline DPGVAE \cite{Yang21} & Edge-ADP & \solidcircle & \solidcircle \\
        \hline DPGGAN \cite{Yang21} & Edge-ADP & \solidcircle & \solidcircle \\
        \hline CAGMDP \cite{Chen20} & (Edge \& Node-Att.)-DP & \solidcircle & \solidcircle \\
        \hline Eli\'{a}\v{s} et al. \cite{Elias20} & Edge-ADP & \emptycircle & \emptycircle \\
        \hline PHDP \cite{Gao19} & Edge-DP & \solidcircle & \emptycircle \\
        \hline LDPGen \cite{Qin17} & Edge-DP & \solidcircle & \solidcircle \\
        \hline TriCycLe \cite{Jorgensen16} & (Edge \& Node-Att.)-DP & \solidcircle & \solidcircle \\
        \hline TmF \cite{Nguyen15} & Edge-DP & \solidcircle & \solidcircle \\
        \hline DER \cite{Chen14} & Dep. Edge-DP & \solidcircle & \solidcircle \\
        \hline
    \end{tabular}
    \caption{Application of \textit{O1} and \textit{O2} for edge-DP mechanism selection in \textit{Scenario 1}. We then evaluate fully-satisfying works. Satisfaction Key: {\small \solidcircle} = Full, {\small \halfcircle} = Partial, {\small \emptycircle} = None.}
    \label{tab:edge-DP-mechanisms}
\end{table}

\subsubsection{O3}
We adopt the following threat model. Given a sensitive input graph $G = (V, E)$, a mechanism $M$ outputs a private version of it, $G'$, under edge-DP, thus protecting edges, i.e., structural information. An adversary has two potential goals: to correctly determine whether some $(u, v) \in V \times V$ is such that $(u, v) \in E$ (edge prediction), and to produce an estimated set of edges $\tilde{E}$ that overlaps substantially with $E$ (edge set reconstruction).  The adversary has no privileged access or background knowledge, only access to $G'$; however, it is known that node identifiers align between $G$ and $G'$.

To evaluate, we perform a link prediction task, reporting accuracy, and apply Azogagh et al.'s GRAND reconstruction attack~\cite{Azogagh25}, reporting relative absolute error, as defined in App. Table~\ref{tab:metrics-definitions}.

\subsubsection{O4}\label{evaluations.scenario-1.o4}
In our utility evaluation, we aim to investigate how different methods preserve structural information. As there are many ways to describe this, we select a large set of descriptive, simulative, and predictive tasks to quantify different aspects of structural preservation. Formal definitions are provided in App.~\ref{formal-definitions}.

\paragraph{Descriptive Metrics}
We group descriptive metrics into local (node-level) and global (graph-level)~\cite{Barabasi16}. For local metrics, we report distributional errors for node degrees, betweenness centralities, closeness centralities, and clustering coefficients, which respectively capture connectivity, information flow, reachability, and cohesion. Using full distributions rather than aggregates provides a nuanced view of structural fidelity. 
Preserving such properties is essential for maintaining the utility of downstream tasks that rely on realistic structural patterns.
We use Wasserstein distance~\cite{Villani09} to measure distribution errors
between input and output graphs, as it captures ordered value shifts, is robust to disjoint data, and maintains interpretability. For the larger datasets (GitHub and Brightkite), computing exact betweenness and closeness centralities is infeasible due to the superquadratic time complexities of their algorithms~\cite{Leskovec06}. Thus, we use the paramaterised approximation algorithms by Riondato and Kornaropoulos~\cite{Riondato14} for betweenness and Cohen et al.~\cite{Cohen14} for closeness, each with their parameters set to ensure errors within $1\%$ of the true values.

Global statistics assess the preservation of global structural properties that underpin many downstream analytical tasks. We report errors in density, harmonic diameter (chosen over standard diameter to remain meaningful for disconnected graphs), assortativity, and modularity. These capture sparsity, reachability, homophily, and community structure, respectively. Preserving such traits is essential for application-oriented utility. For example, sparsity, small-world characteristics, and ``birds-of-a-feather'' community behaviour are expected in social graphs~\cite{Watts98, McPherson01}. We report absolute errors for all global metrics, except for harmonic diameter, where relative error is used due to its scale-dependent nature.

We also report the Adjusted Rand Index (ARI)~\cite{Hubert85}, which measures the chance of agreement between clusterings. This assesses the preservation (utility) of clusterings within the original and private graphs. As the datasets lack community labels, we obtain partitions detected by the Louvain method~\cite{Blondel08}.

\paragraph{Simulative Tasks}
We simulate influence maximisation to measure the preservation of realistic diffusion processes. Relevant applications include information dissemination~\cite{Singh19} and behavioural modelling~\cite{Zareie23}. For each graph, we run \num{1000} simulations using the TIM+ algorithm~\cite{Tang14} under the Independent Cascade diffusion model~\cite{Kempe03}, starting from $1\%$ of the nodes. We report the absolute error in the percentage of influenced nodes to quantify how closely the private graph replicates the dynamics of the original.

\paragraph{Predictive Tasks}
To quantify the utility of predictive analyses, we compare the performance of ranking- and outcome-inference tasks across non-private and private graphs. Predictive task performance is a direct proxy for utility in many real-world applications. For recommendation tasks, we report the Normalised Discounted Cumulative Gain (NDCG)~\cite{Jarvelin02}, which accounts for both the relevance and ranking. Node scores are computed with PageRank~\cite{Page99}, a widely adopted metric balancing influence and communicability. 

We evaluate utility on two learning tasks: link prediction~\cite{Zhang18} and node classification~\cite{Kipf17}. For private graphs, we train a two-layer Graph Convolutional Network (GCN)~\cite{Kipf17} and report absolute errors in the AUROC for an edge-prediction task on datasets with labelled nodes (LastFM and GitHub). For node classification, we use another two-layer GCN with feature augmentation generated from a Node2Vec model, using all recommended parameters from the original publication~\cite{Grover16}. We report the absolute error in the F1-score. These metrics were selected for their robustness under class imbalance, a common property of real-world graphs~\cite{Barabasi16}.

Both GCNs use a hidden layer dimension of \num{256}, a $25\%$ dropout rate, and the Adam optimiser~\cite{Kingma14} with a learning rate of \num{0.01}. These hyperparameters and architectures were selected as standard, broadly applicable defaults commonly used in the literature~\cite{Kipf17}. Our goal is not to optimise model performance but to ensure a consistent setup for comparing the relative impact of graph privatisation methods on downstream learning tasks. We use a \num{80}/\num{10}/\num{10} split for training, validation, and testing.  To accurately evaluate utility preservation, the same test set used for the unprotected graph is applied to the private graph, isolating the effect of privatisation.
This ensures that any observed differences in utility are due to privatisation, rather than random variations in test set selection. 

\subsubsection{Utility Results and Discussion}\label{evaluations.scenario-1.utility-results}
\begin{figure*}[htbp]
    \centering
    \includegraphics[width=0.498\linewidth]{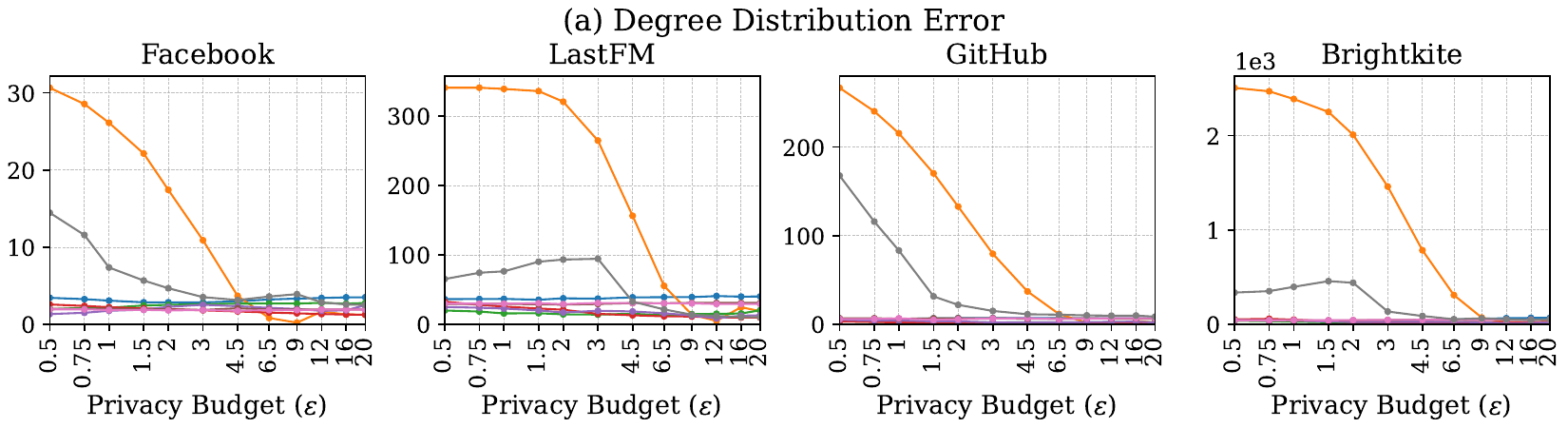}
    \hfill
    \includegraphics[width=0.498\linewidth]{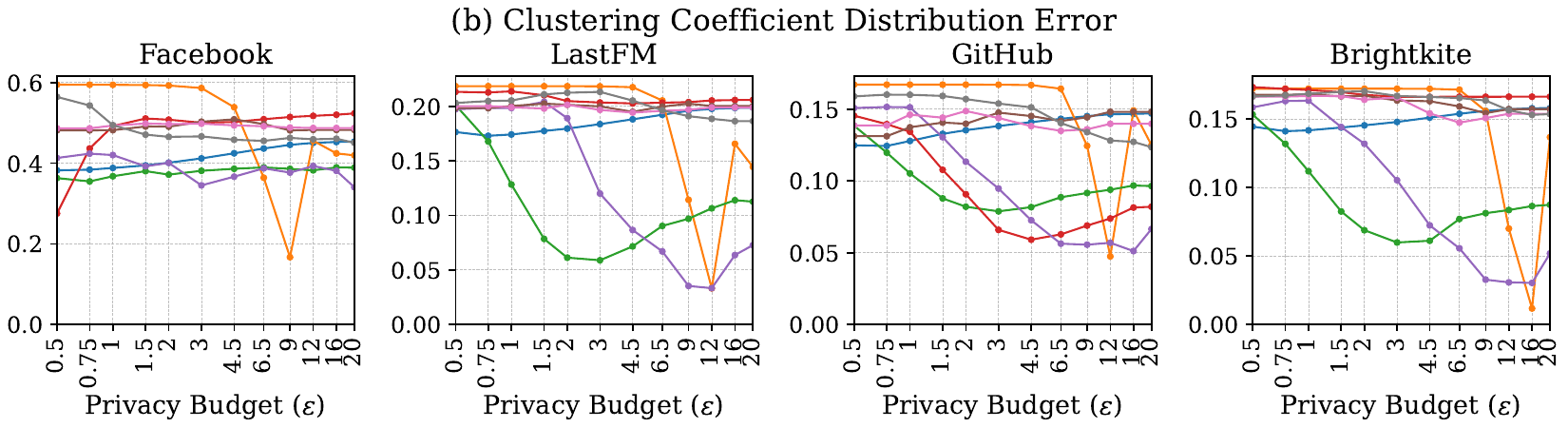}
    \includegraphics[width=0.498\linewidth]{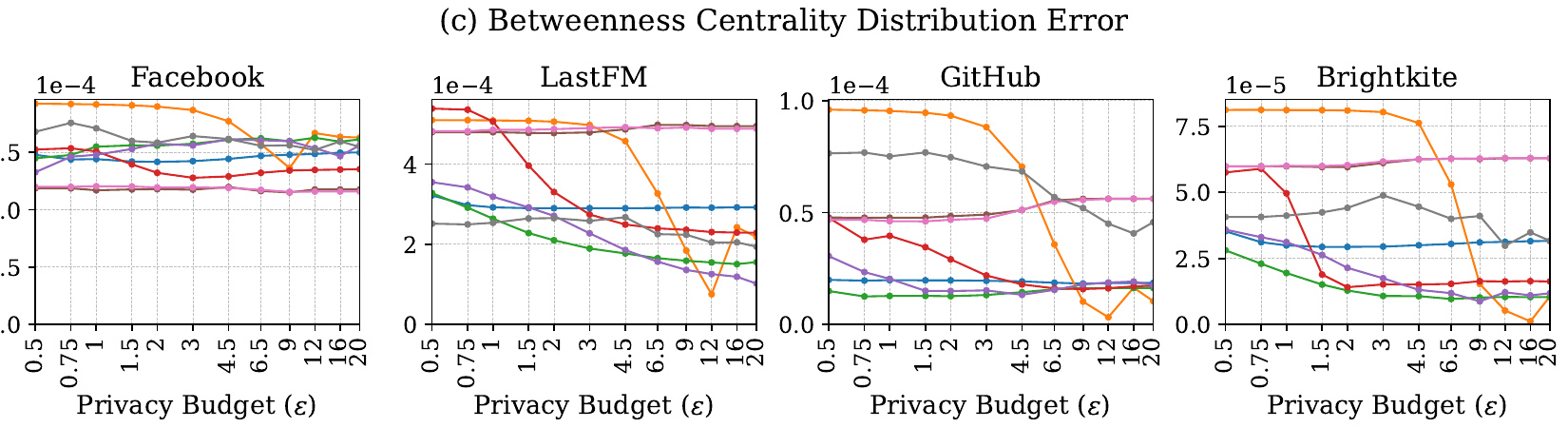}
    \hfill
    \includegraphics[width=0.498\linewidth]{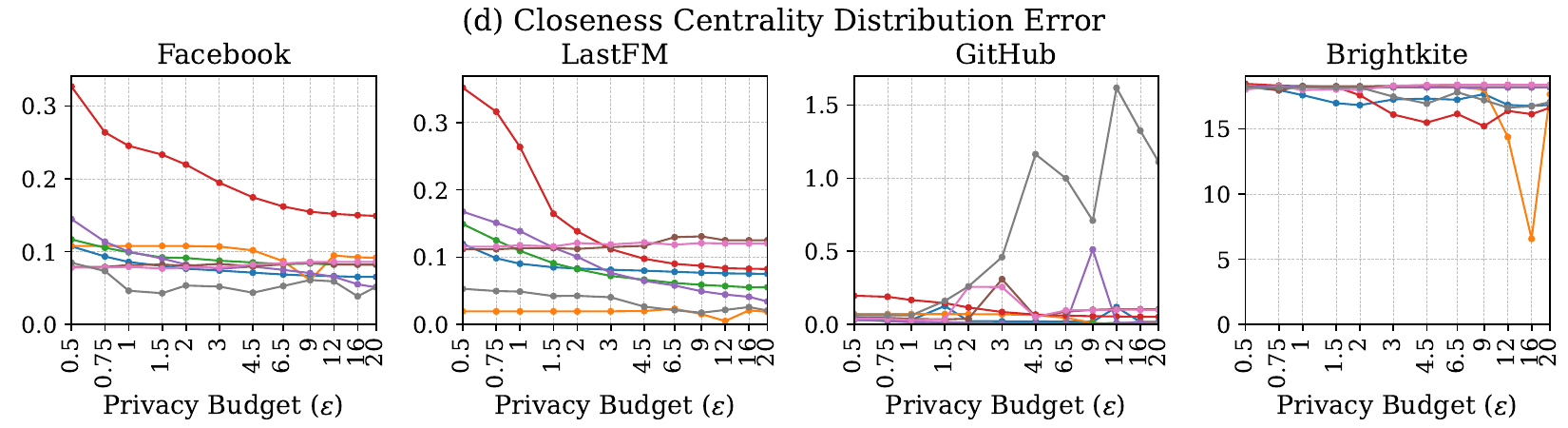}
    \includegraphics[width=0.498\linewidth]{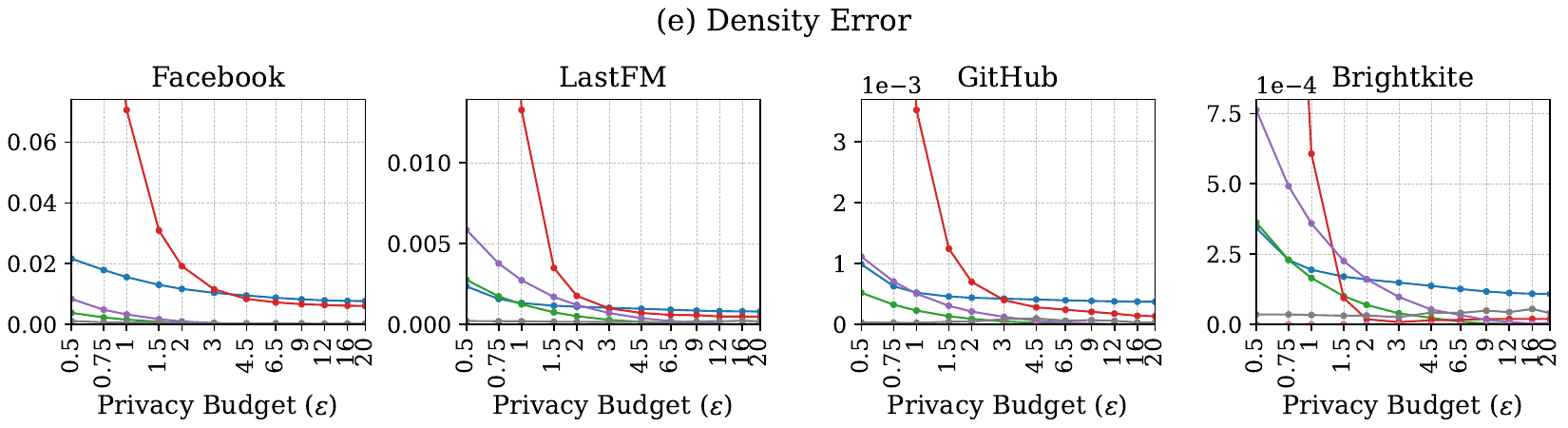}
    \hfill
    \includegraphics[width=0.498\linewidth]{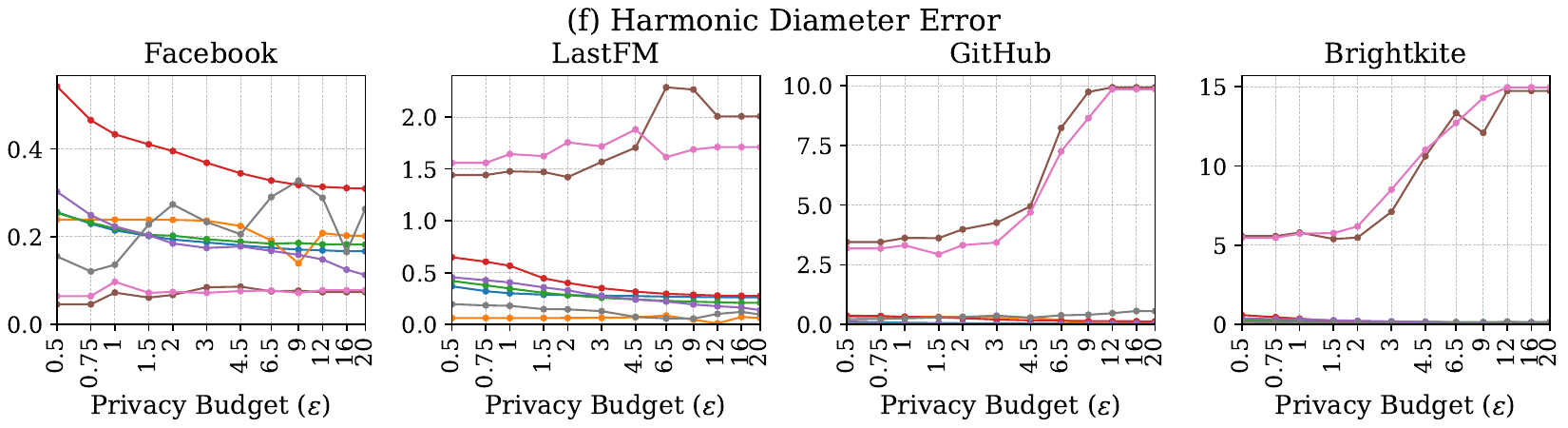}
    \includegraphics[width=0.498\linewidth]{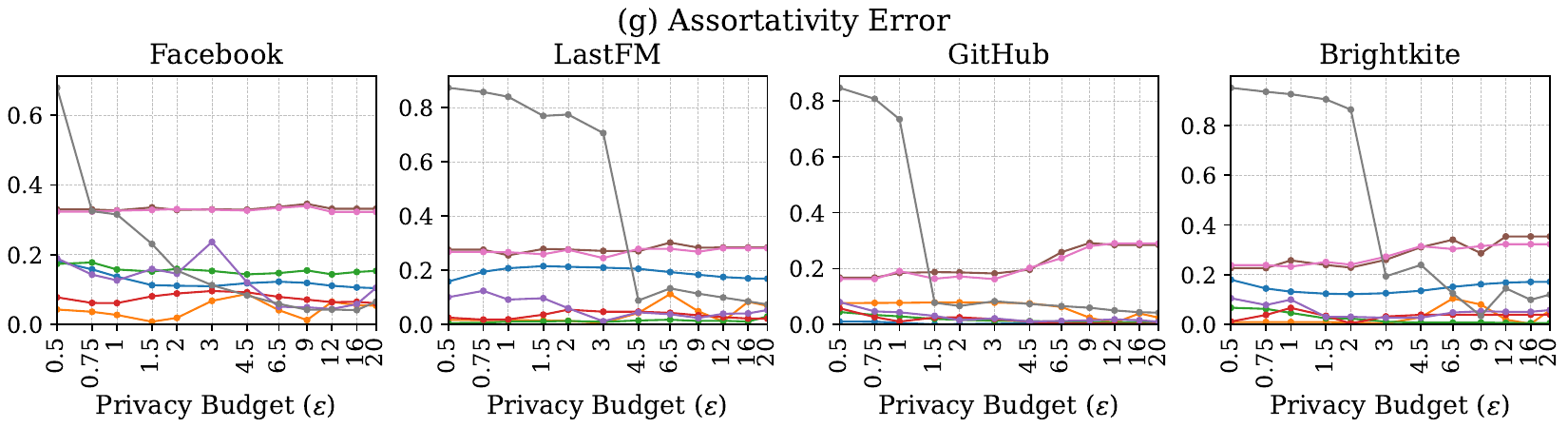}
    \hfill
    \includegraphics[width=0.498\linewidth]{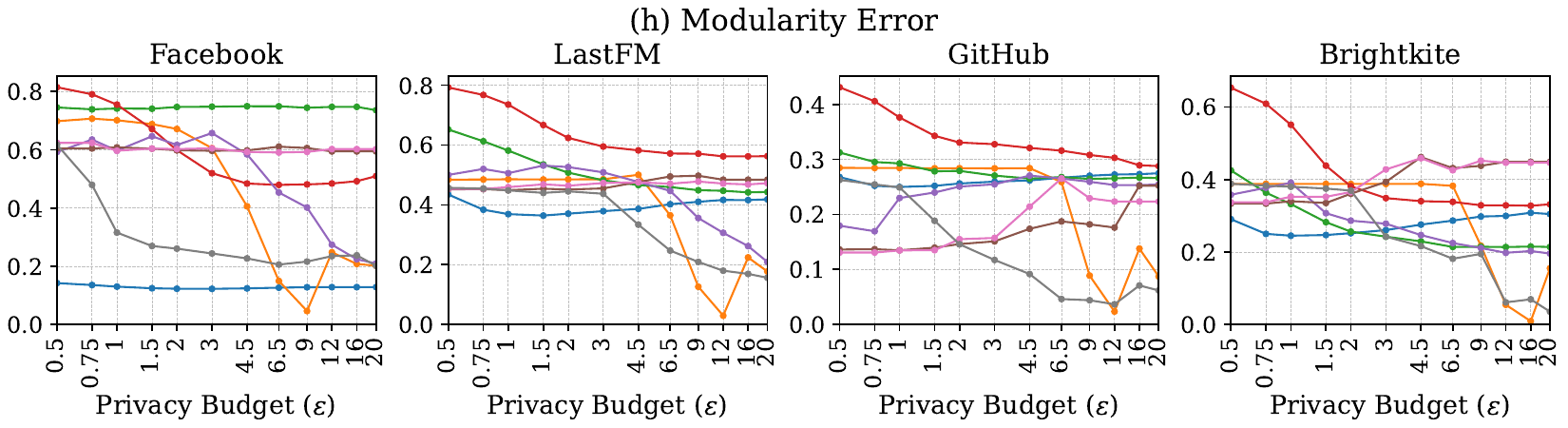}
    \includegraphics[width=0.498\linewidth]{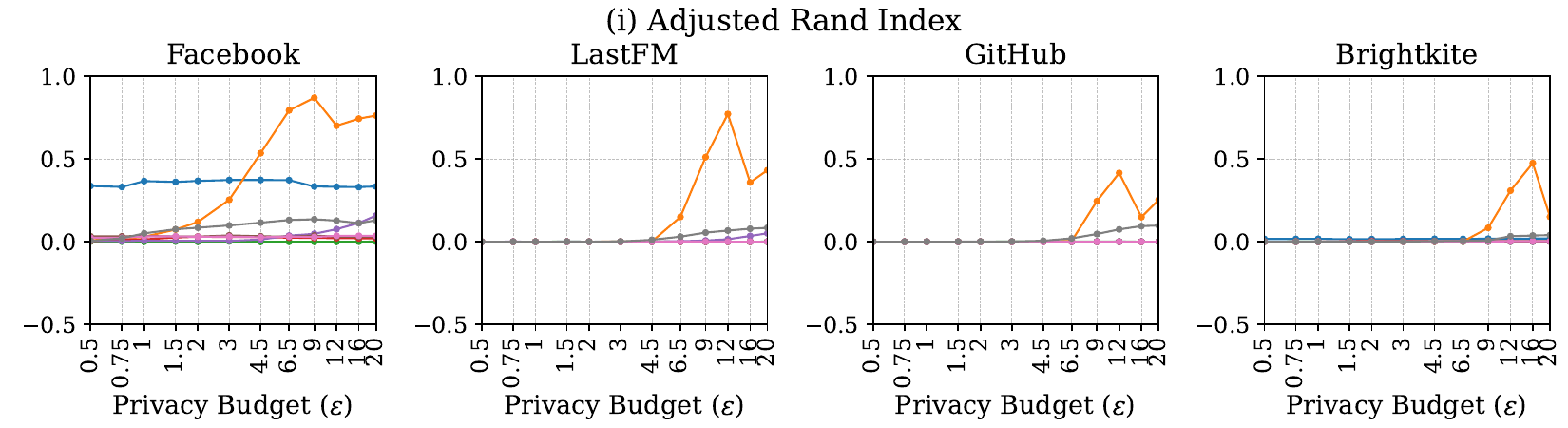}
    \hfill
    \includegraphics[width=0.498\linewidth]{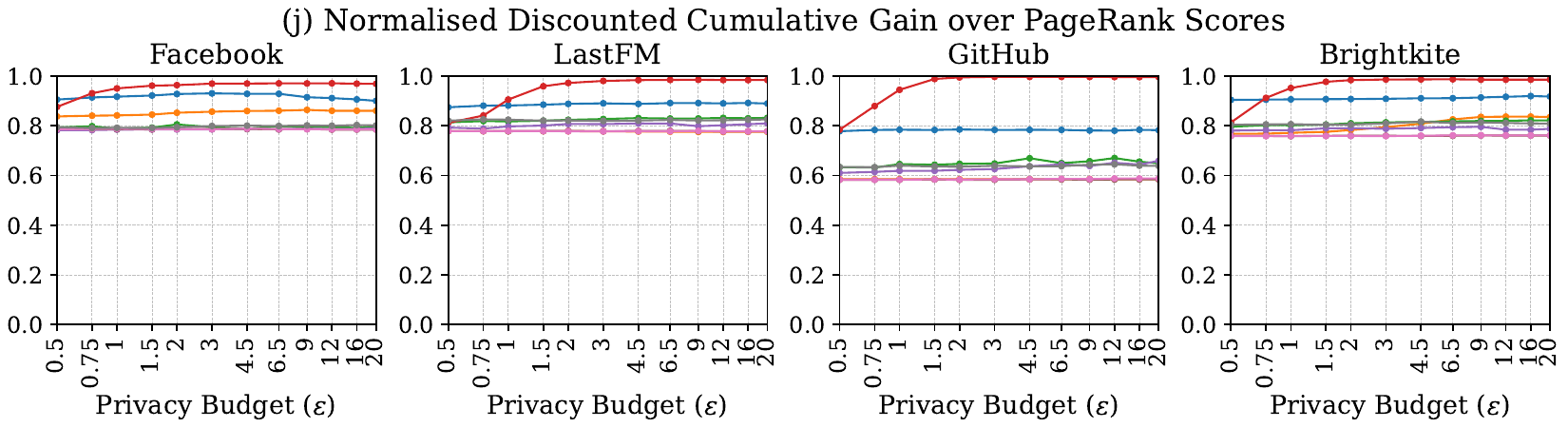}
    \includegraphics[width=0.498\linewidth]{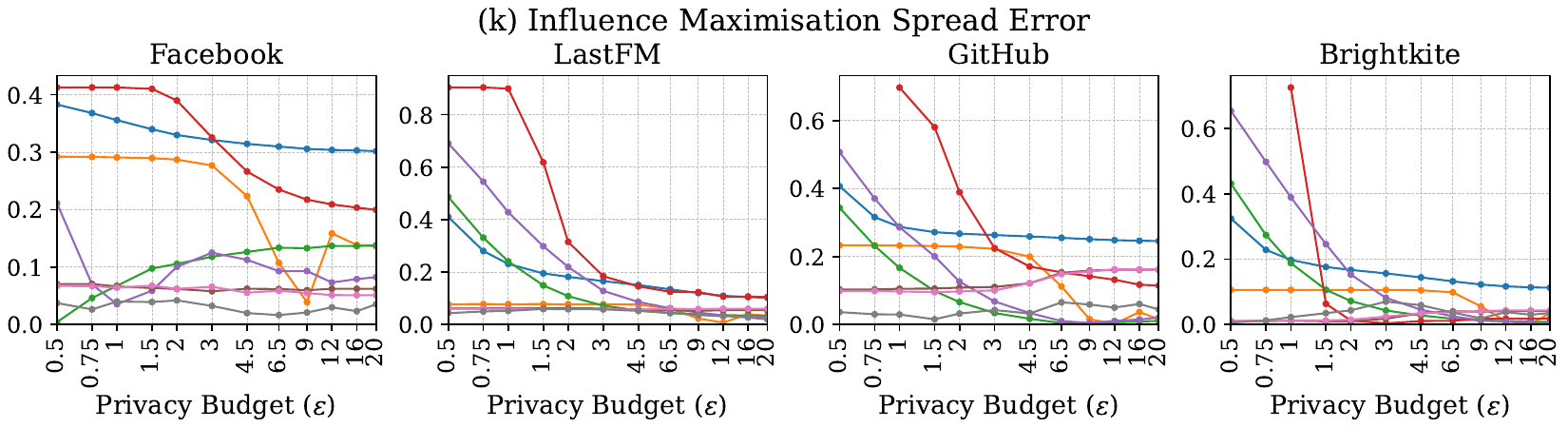}
    \hfill
    \includegraphics[width=0.498\linewidth]{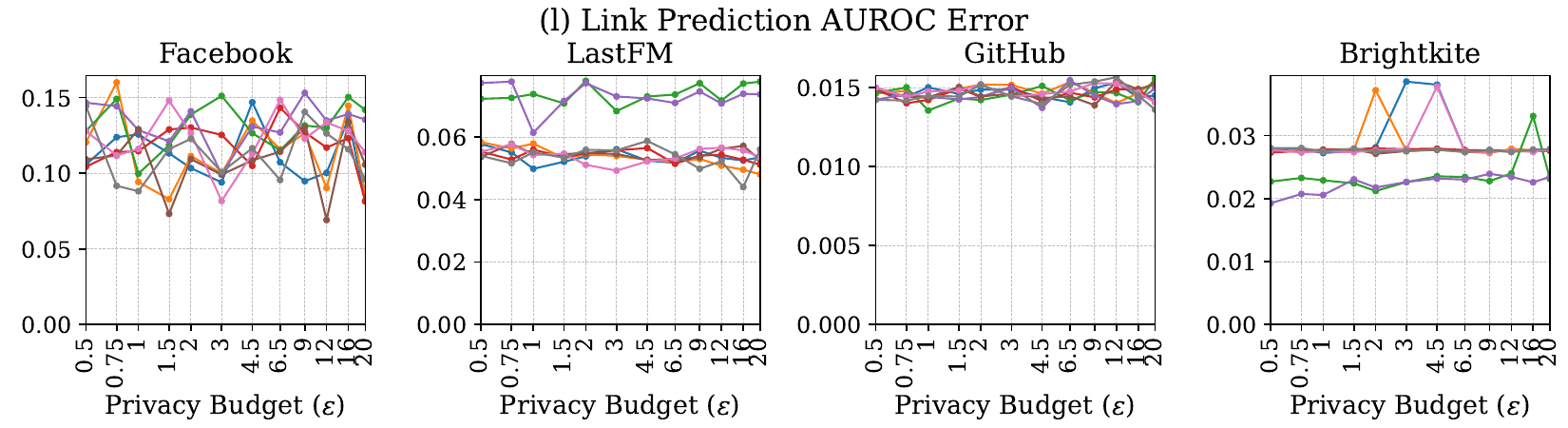}
    \includegraphics[width=0.25\linewidth]{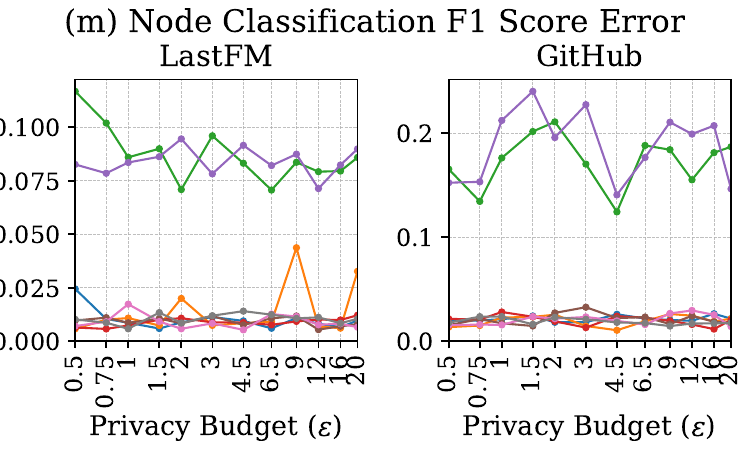}
    \hfill
    \raisebox{0.4cm}{\includegraphics[width=0.21\linewidth]{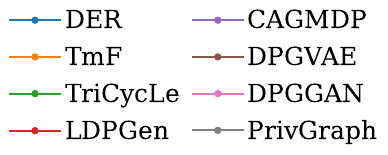}}
    \hfill
    \begin{minipage}[b]{0.475\linewidth}
        \caption{Results for Scenario 1. Distribution errors are quantified with the Wasserstein distance; harmonic diameter with relative error; and remaining error-based metrics with absolute error. All values are averaged over 10 independent trials. Each metric is assigned an alphabetical identifier.}
        \label{fig:scenario-1-results}
    \end{minipage}
\end{figure*}

We present results as plots (Fig.~\ref{fig:scenario-1-results}) due to the multidimensional nature of our measurements. In our repository, we provide numerical, tabular results and further plots. 
During the evaluation, we found that LDPGen significantly overestimated graph density in smaller privacy budgets, rendering our influence maximisation simulations computationally infeasible. Thus, we excluded their evaluation for privacy budgets $\epsilon \in [0.5, 0.75]$ for the GitHub and Brightkite datasets.

\paragraph{Descriptive Analyses}
Our descriptive graph metrics capture topological structure, providing insights into reachability, connectivity, node importance, clusterings, and community structure. Their preservation is quantified by measuring errors between metrics on non-private and private graphs. Lower errors imply greater utility.

Reachability and connectivity are quantified with density, harmonic diameter, and degree distributions. Most methods replicate graph density with reasonable accuracy; however, smaller $\epsilon$ tend to produce overly dense graphs, as shown in Fig.~\ref{fig:scenario-1-results}(e). This is a known trait of edge-DP methods, where false edges are added for indistinguishability~\cite{Zhang20t}. This inflates connectivity and reachability, most notably in LDPGen due to its stricter local-DP guarantees. Harmonic diameters (Fig.~\ref{fig:scenario-1-results}(f)) are generally well-preserved; however, the deep learning methods (DPGGAN and DPGVAE) incur higher errors on larger datasets, implying a bias towards underestimating connectivity. Since these methods accurately estimate density, inflated harmonic diameters suggest reachability issues. Degree distributions are approximated well by most methods (Fig.~\ref{fig:scenario-1-results}(a)), though TmF and PrivGraph underperform under smaller $\epsilon$. A closer inspection of their respective degree distributions (available in our repository) indicates a tendency to overestimate the proportion of lower-degree nodes, leading to an underestimation of connectivity.

Node importance, underpinning many downstream tasks such as influence maximisation, is measured using centrality metrics that correlate with connectivity. We observe similarities between degree and betweenness (Fig.~\ref{fig:scenario-1-results}(c)) and closeness centrality (Fig.~\ref{fig:scenario-1-results}(d)) distributions, confirming the link between local connectivity and shortest-path structures. Edge-and-node-attribute-DP methods, CA-GMDP and TriCycLe, preserve betweenness most consistently. Generally, methods replicate closeness less accurately, likely because it relies on global reachability. Larger errors were observed at smaller $\epsilon$ for LDPGen, likely due to inflated density, and PrivGraph, whose elevated harmonic diameter errors imply reachability limitations.

For most associative and community-based metrics, most methods exhibit minimal sensitivity to $\epsilon$. We observe in Fig.~\ref{fig:scenario-1-results}(b) that clustering coefficient distributions are generally better preserved in larger datasets, with TmF, CAGMDP, and TriCycLe performing most competitively. By contrast, PrivGraph underestimates assortativity (Fig.~\ref{fig:scenario-1-results}(g)) for smaller $\epsilon$, failing to capture homophily, the tendency of similar nodes to connect, resulting in larger errors than other methods. Across all methods, modularity is underestimated (Fig.~\ref{fig:scenario-1-results}(h)), suggesting a shared weakness in replicating community structures. Nevertheless, PrivGraph, TmF, and CAGMDP showed improvement with larger $\epsilon$. Lower assortativity and reduced modularity preservation are expected to result in poorer community preservation, confirmed by ARI results: most methods yield near-zero ARI across all $\epsilon$, shown in Fig.~\ref{fig:scenario-1-results}(i), implying randomised community labels. Only TmF and DER (for the Facebook dataset only) demonstrate preservation at moderate to large $\epsilon$.

In summary, most methods preserve basic structural properties reasonably well; only a subset, particularly CAGMDP, TriCycLe, and TmF, consistently achieve stronger utility for descriptive analyses. More nuanced properties, such as centralities and community structures, vary significantly across methods. To maximise utility, practitioners should prioritise the preservation of specific structural features most relevant to their intended tasks. 

\paragraph{Simulative Analyses}
Influence maximisation was measured by spread error, with results shown in Fig.~\ref{fig:scenario-1-results}(k). Since this task depends heavily on connectivity, results strongly correlate with the connectivity metrics. Higher connectivity accelerates diffusion, leading to larger spreads—most evident in methods that overestimate density, such as LDPGen, DER, CAGMDP, and TriCycLe. Interestingly, TmF, despite accurate density preservation, shows high spread error, likely due to poor preservation of node roles along paths. Conversely, DPGGAN and DPGVAE exhibit reduced spread accuracy, consistent with inflated harmonic diameters and longer paths. 

These observations highlight how connectivity properties affect influence propagation: higher degrees improve hub placement, while larger harmonic diameters suggest slower diffusion due to longer average paths between nodes. Consequently, methods that best preserve utility for influence maximisation are generally those that most effectively preserve connectivity. Under a smaller $\epsilon$, PrivGraph is the most consistent, while under a larger $\epsilon$, TmF, TriCycLe, DPGGAN, and DPGVAE become viable options.

\paragraph{Predictive Tasks}
The PageRank NDCG results (Fig.~\ref{fig:scenario-1-results}(j)) indicate decent utility for PageRank-related ranking and recommendation tasks. Most methods preserve node rankings well, with LDPGen achieving nearly perfect alignment under moderate and larger $\epsilon$.

Errors in link prediction and node classification performance are shown in Figs.~\ref{fig:scenario-1-results}(l) and \ref{fig:scenario-1-results}(m), respectively. Across both tasks, most methods perform similarly, apart from the edge-and-node-attribute-DP methods, CAGMDP and TriCycLe. Their lower utility is explained by the added noise required to privatise semantic attributes, which directly impacts feature quality. Despite this, predictive performance error remains generally low across methods.

We measure predictive performance error rather than raw accuracy to assess utility preservation. This highlights a key tension: strong predictive utility can conflict with privacy guarantees. For example, accurate link prediction on edge-DP graphs constitutes a reconstruction attack, since predicted edges replicate those of the original sensitive graph, undermining the edge-DP guarantee. Similarly, precise node classification implies effective recovery of node attributes (i.e., target classes), increasing susceptibility to node re-identification~\cite{Horawalavithana19}, opposing the goal of node-DP.

We formalise this trade-off using the hypothesis-testing interpretation of DP by Kairouz et al.~\cite{Kairouz15}. Let $A$ be an $\epsilon$-DP mechanism and $F: A(\cdot) \mapsto \{0, 1\}$ be a binary predictor on the privacy target (e.g. $A$ may be an edge-DP mechanism and $F$ a link predictor). Then, we have $\alpha \le e^\epsilon \beta$, where $\alpha = \text{Pr}[F(A(X)) = 1]$ and $\beta = \text{Pr}[F(A(X')) = 1]$. With a uniform prior, the accuracy of $F$ is $F_\text{acc} = \frac{1}{2}\alpha + \frac{1}{2}(1-\beta) = \frac{1}{2} + \frac{\alpha - \beta}{2}$. We also have $\alpha - \beta \le (e^\epsilon - 1)\beta \le e^\epsilon - 1$. Therefore, the prediction advantage over random guessing is $F_\text{adv} = F_\text{acc} - \frac{1}{2} = \frac{\alpha - \beta}{2} \le \frac{e^\epsilon - 1}{2}$. Then, to achieve $F_\text{adv} > 0$, the required privacy budget is lower bounded by $\epsilon \ge \ln(2F_\text{adv} + 1)$. This reflects the trade-off between privacy and prediction accuracy when the privacy and prediction targets align, such as in edge-DP versus link prediction, or node-DP versus node re-identification.
 
Thus, when downstream predictive tasks are known in advance, it is crucial to assess whether privacy and utility are fundamentally opposed in that setting, and whether compromises in privacy budgets, notions, and granularity are necessary.

\subsubsection{Empirical Privacy Results and Discussion}\label{evaluations.scenario-1.empirical-privacy}
\begin{figure}[tbp]
    \centering
    \hspace*{0.02\textwidth}
    \includegraphics[clip, trim=0cm 0cm 21cm 1.2cm, width=0.2\textwidth]{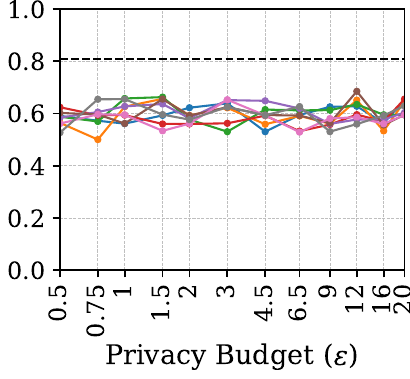}
    \hfill
    \raisebox{0.75cm}{\includegraphics[width=0.21\textwidth]{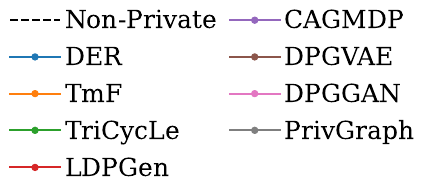}\hspace*{0.02\textwidth}}
    \caption{Link prediction accuracy against edge-DP methods, evaluated on the Facebook dataset.}
    \label{fig:edge-dp-link-pred-accuracy-facebook}
\end{figure}

Results for the Facebook dataset are shown in Fig.~\ref{fig:edge-dp-link-pred-accuracy-facebook} for the link prediction task and Table~\ref{tab:edge-dp-reconstruction-facebook} for the reconstruction attack. Variability across methods and $\epsilon$ is minimal: link prediction accuracy remains around $60\%$, and reconstruction attacks consistently reduce relative absolute error to about $78\%$, cutting error by more than half for most methods.

While these tasks do not achieve optimal success, they recover a significant amount of sensitive information (the edges, in this case). The consistent success across methods and budgets suggests a shared vulnerability. To identify this, we refer to our systemisation diagram (Fig.~\ref{systemisation-diagram}). 
Starting from the bottom, we evaluated link prediction, which simultaneously serves as a membership inference attack under our threat model (as discussed in Sec.~\ref{evaluations.scenario-1.utility-results}), plus a reconstruction attack.
Tracing the relevant lines up the right-hand side, we identify a large set of possible vulnerabilities; however, the one that is critically shared amongst all methods is the data independence assumption issue. As shown in~\cite{Kifer11} and noted in Sec.~\ref{systemisation.approaches.privacy-mechanism}, edges are rarely independent in practice, yet edge-DP mechanisms implicitly assume so.

The observed consistency of the empirical attacks despite varying $\epsilon$ indicates that while the DP mechanisms calibrate noise to individual edge-level sensitivity, the attacks exploit global and multi-hop structural correlations. In the large, sparse graphs (typical of social networks~\cite{Barabasi16}) evaluated here, such structural patterns have higher signal-to-noise ratios; thus, varying $\epsilon$ is not guaranteed to significantly affect higher-order features leveraged by the attack. These results indicate a substantial empirical privacy loss, an essential consideration for practitioners.

\begin{table}
  \small
  \centering
  \begin{tabular}{c|ccc|ccc}
    \hline \textbf{Method} & \multicolumn{3}{c|}{\textbf{Private Graph}} & \multicolumn{3}{c}{\textbf{Reconstruction}} \\
    \hline & $\epsilon = 1$ & $\epsilon = 3$ & $\epsilon = 9$ & $\epsilon = 1$ & $\epsilon = 3$ & $\epsilon = 9$ \\
    \hline DER & 165.046 & 152.772 & 149.11 & 78.641 & 78.741 & 78.719 \\
    \hline TmF & 140.268 & 139.625 & 136.42 & 78.594 & 78.59 & 78.683 \\
    \hline TriCycLe & 144.936 & 140.939 & 139.82 & 78.779 & 78.668 & 78.656 \\
    \hline LDPGen & 278.242 & 151.122 & 141.796 & 78.548 & 78.765 & 78.825 \\
    \hline CAGMDP & 150.165 & 141.663 & 139.61 & 78.828 & 78.648 & 78.593 \\
    \hline DPGVAE & 140.662 & 140.638 & 140.629 & 78.528 & 78.731 & 78.7 \\
    \hline DPGGAN & 140.626 & 140.689 & 140.706 & 78.856 & 78.63 & 78.729 \\
    \hline PrivGraph & 133.534 & 136.287 & 136.616 & 78.607 & 78.64 & 78.735 \\
    \hline
  \end{tabular}
  \caption{Relative absolute error of the reconstruction attack~\cite{Azogagh25} against edge-DP methods on the Facebook dataset.}%
  \label{tab:edge-dp-reconstruction-facebook}
\end{table}

\subsection{Example Scenario 2: Node Privacy}\label{evaluations.scenario-2}
For the second scenario, we assume a similar role to the first, but focus on node privacy rather than edges. For \textit{O1}, we use the same datasets and accept the same guarantees as \textit{Scenario 1}, but define our privacy target as the nodes. For \textit{O2}, we use the same machine configuration and similarly express these first two objectives in Table~\ref{tab:node-dp-mechanisms}. We evaluate all node-DP mechanisms from our taxonomy in Table~\ref{tab:sok}, noting that PrivCom~\cite{Zhang20} is only partially efficient due to its super-quadratic time complexity, limiting its evaluation to the two smaller datasets.

\begin{table}
    \small
    \centering
    \begin{tabular}{c|ccc}
        \hline \textbf{Work} & \textbf{Privacy Mechanism} & \textbf{Efficient} & \textbf{Reproducible} \\
        \hline PrivDPR \cite{Zhang25} & Node-ADP & \solidcircle & \solidcircle \\
        \hline $\pi_v$, Jian et al. \cite{Jian23} & Node-DP & \solidcircle & \solidcircle \\
        \hline $\pi_e$, Jian et al. \cite{Jian23} & Node-ADP & \solidcircle & \solidcircle \\
        \hline PrivCom \cite {Zhang20} & Node-ADP & \halfcircle & \solidcircle \\
        \hline
    \end{tabular}
    \caption{Application of \textit{O1} and \textit{O2} for node-DP mechanism selection in \textit{Scenario 2}. Satisfaction Key: {\small \solidcircle} = Full, {\small \halfcircle} = Partial.}
    \label{tab:node-dp-mechanisms}
\end{table}

\subsubsection{O3}\label{evaluations.scenario-2.o3}
We adopt the following threat model. Given a sensitive input graph $G$, a mechanism $M$ outputs a private version of a graph, $G' = (V', E')$, under node-DP, thus protecting against the inference of any node's presence in the graph, i.e., node identification. An adversary attempts to de-anonymise (i.e., re-identify) the nodes in the graph by producing a mapping $F: V' \mapsto V$ of nodes from the private graph $G'$ to those of the sensitive graph $G$ (node de-anonymisation). The adversary has no privileged access or background knowledge, only access to $G'$. They also have no partial mapping of pre-known seeds (anchor pairings) from $V'$ to $V$, and thus must conduct a seed-free de-anonymisation attack. 

We evaluate node-DP methods on the seed-free de-anonymisation attack by Pedarsani et al.~\cite{Pedarsani13}. However, as noted in Sec.~\ref{systemisation.approaches.privacy-mechanism} (\textit{L3} of our systemisation), edges or nodes may be perturbed to achieve node-DP, meaning some methods do not maintain consistent node mappings, preventing ground-truth evaluation of mapping accuracy. Thus, we adopt two popular ground-truth-agnostic metrics from graph de-anonymisation~\cite{Narayanan09} and alignment literature~\cite{Corominas23, El-Kebir15}: edge correctness, and the symmetric substructure score.

\subsubsection{O4} 
We use the same local and global graph metrics described in Sec.~\ref{evaluations.scenario-1.o4}, with additional relative error for the number of nodes and edges due to variability in graph sizes. For the same reason, we excluded partition-based descriptive metrics and the NDCG recommendation ranking task, as these require the node dimensions to be similar. We further excluded the simulation and remaining predictive tasks due to PrivCom's tendency to produce overly dense graphs, as well as the methods by Jian et al., which produced graphs that were too small, rendering these comparisons unsound. Due to PrivCom's inefficiency, evaluation was feasible only on the two smaller datasets, Facebook and LastFM. 

To summarise, we report the errors in the number of nodes, number of edges, density, harmonic diameter, assortativity, and modularity. The errors in the number of nodes, the number of edges, and the harmonic diameter are quantified using relative errors, given their scale-dependent nature. All others are quantified with absolute error. We also use the Wasserstein distance to quantify errors in the distributions of degrees, clustering coefficients, and centralities (betweenness and closeness).

\subsubsection{Utility Results and Discussion}
\begin{figure*}[htbp]
    \centering
    \includegraphics[width=0.498\linewidth]{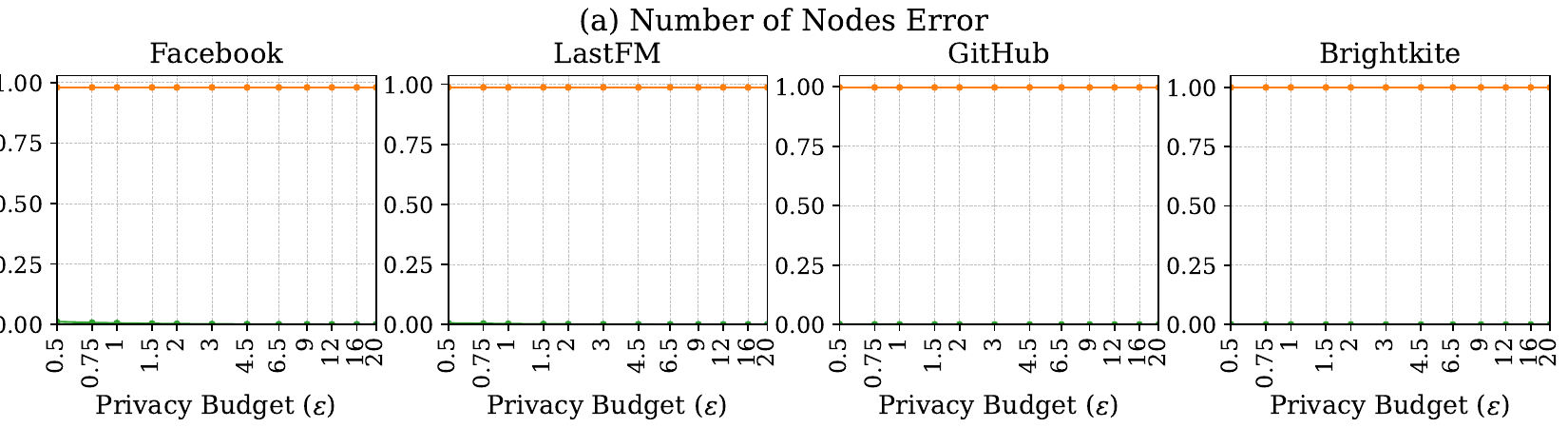}
    \hfill
    \includegraphics[width=0.498\linewidth]{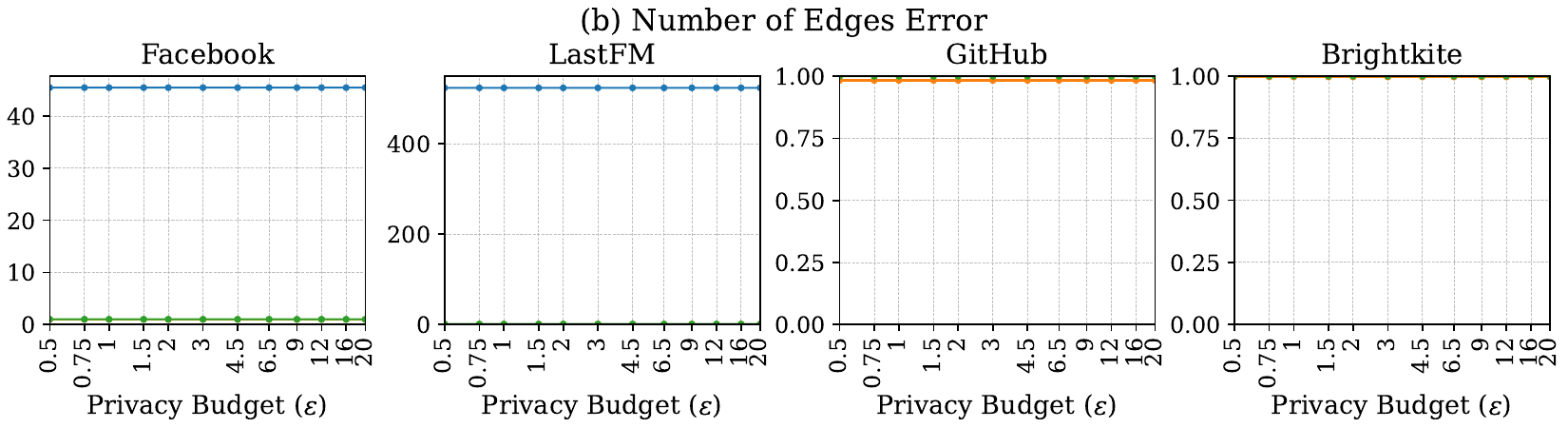}
    \includegraphics[width=0.498\linewidth]{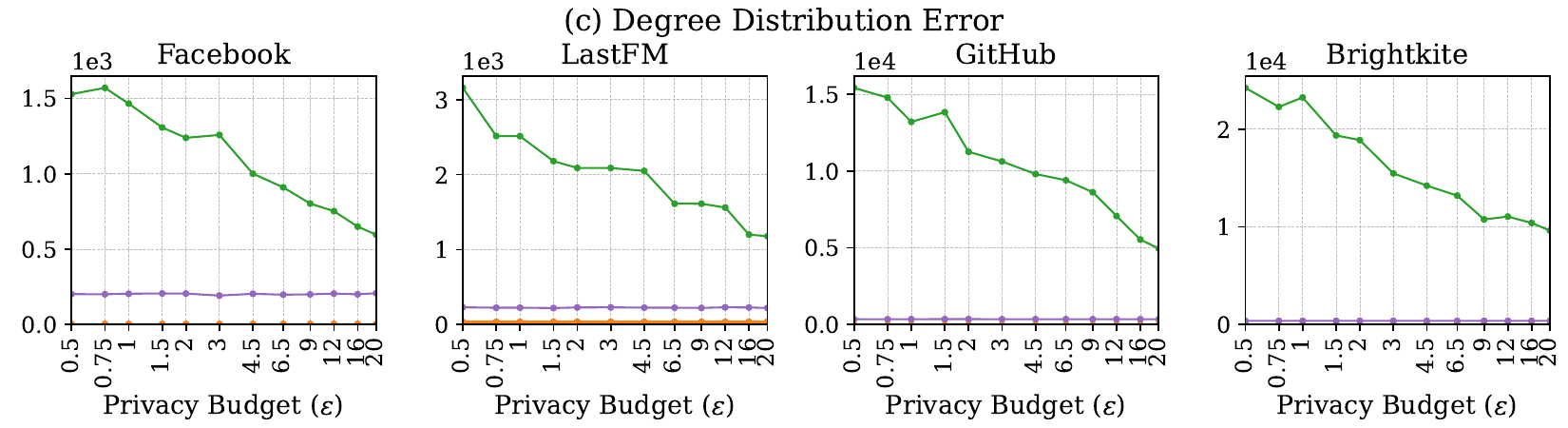}
    \hfill
    \includegraphics[width=0.498\linewidth]{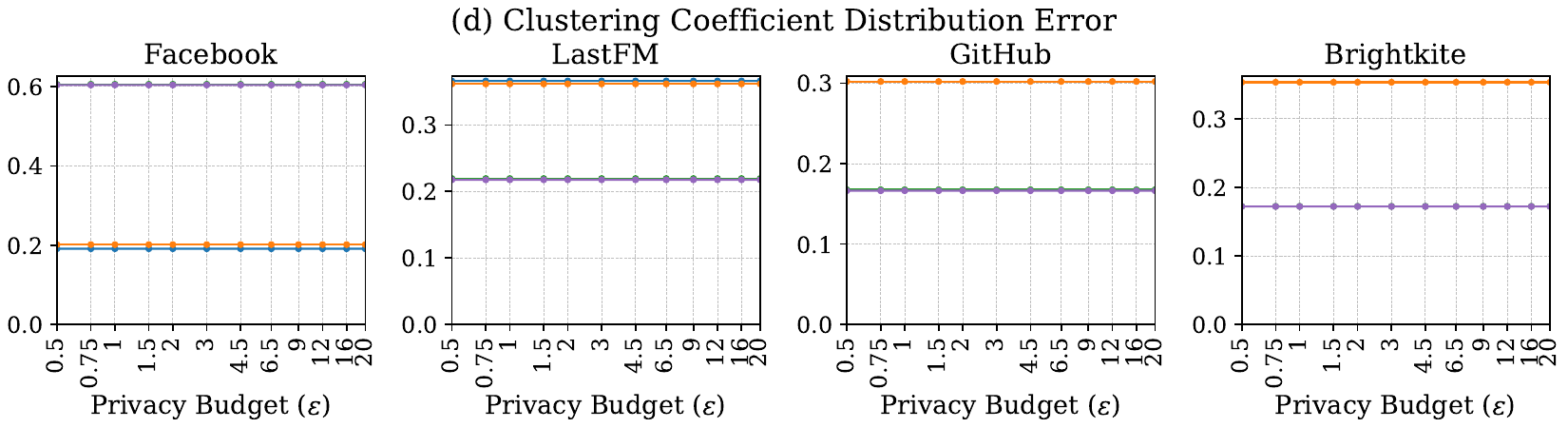}
    \includegraphics[width=0.498\linewidth]{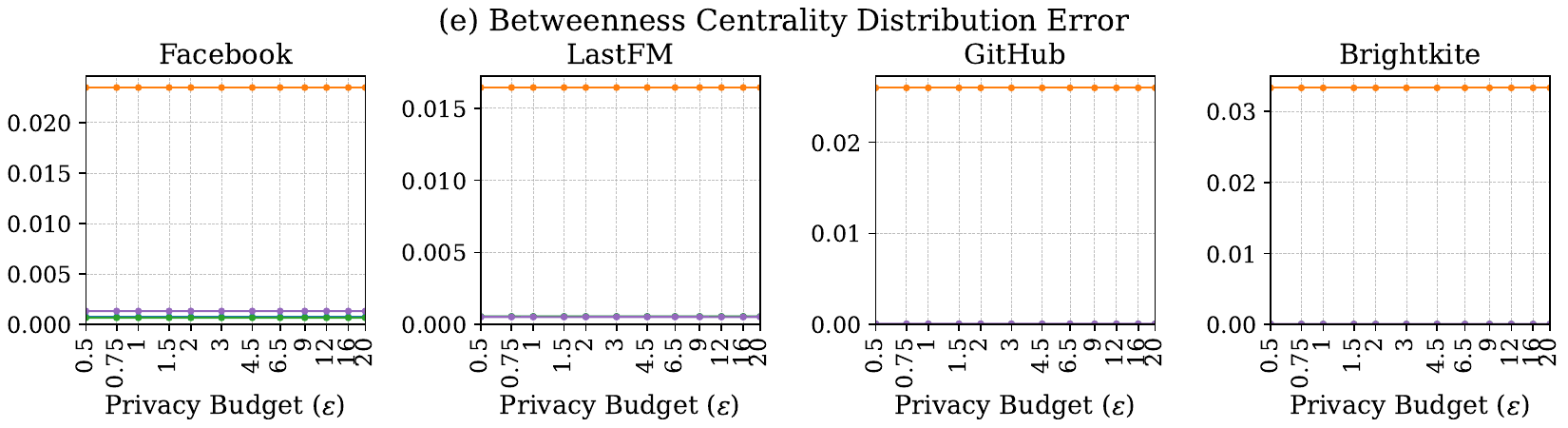}
    \hfill
    \includegraphics[width=0.498\linewidth]{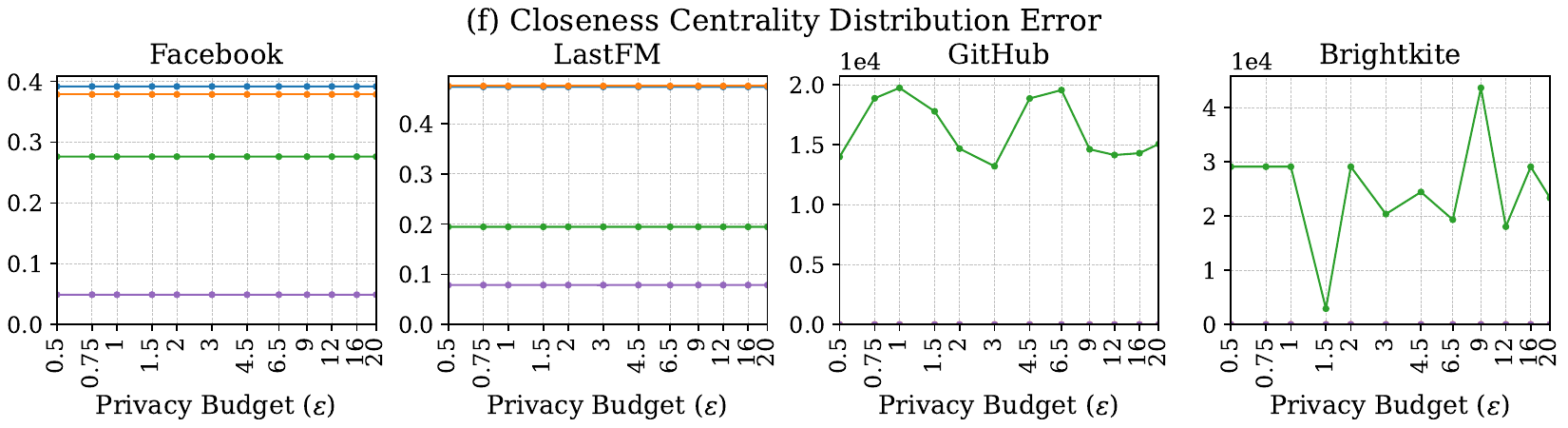}
    \includegraphics[width=0.498\linewidth]{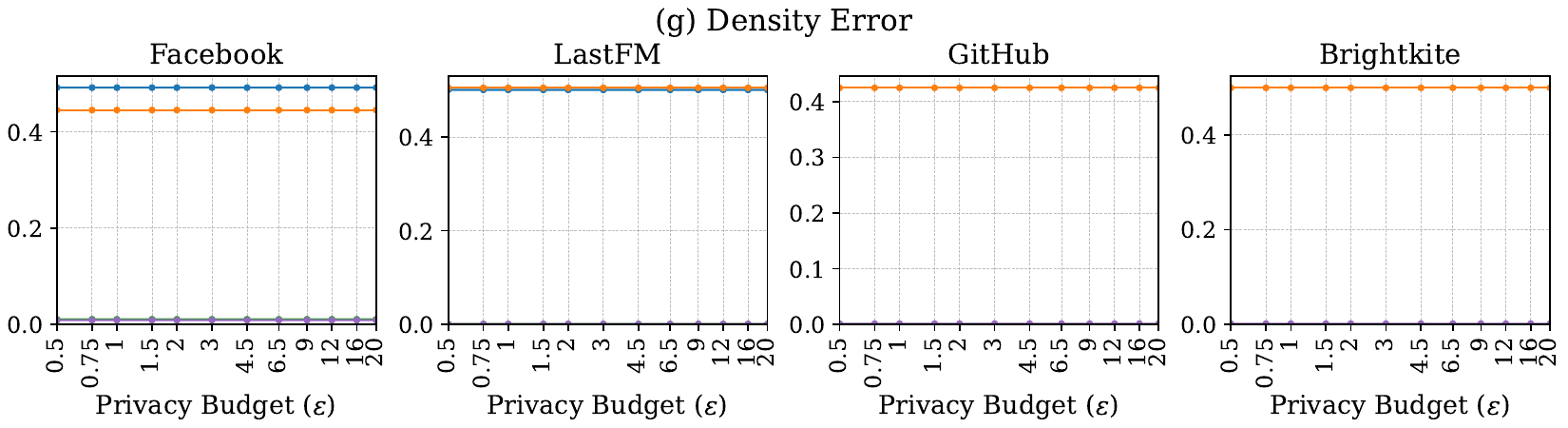}
    \hfill
    \includegraphics[width=0.498\linewidth]{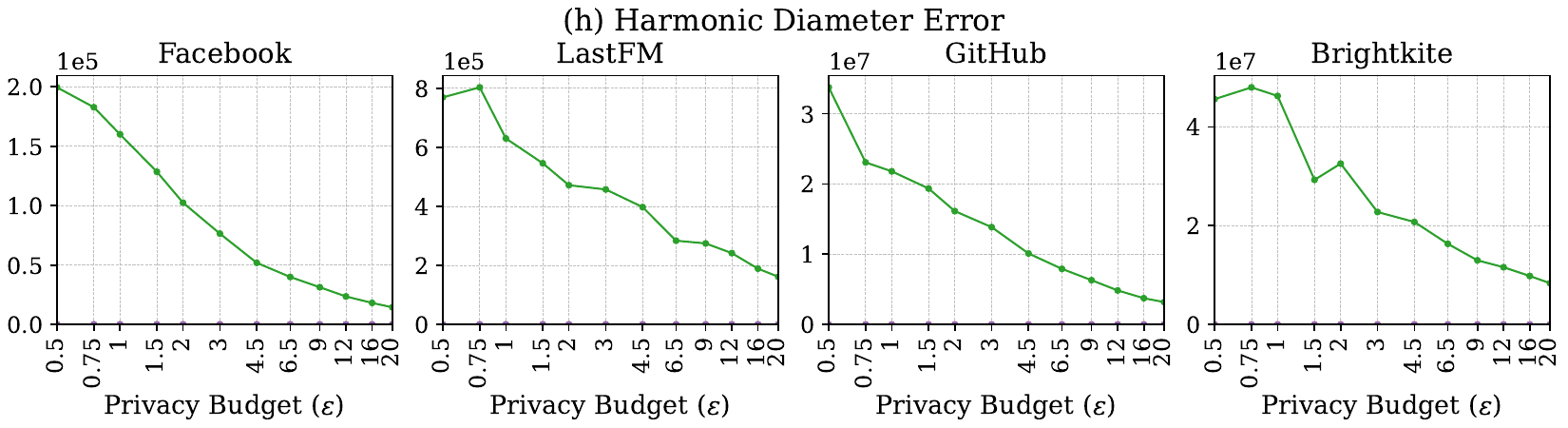}
    \includegraphics[width=0.498\linewidth]{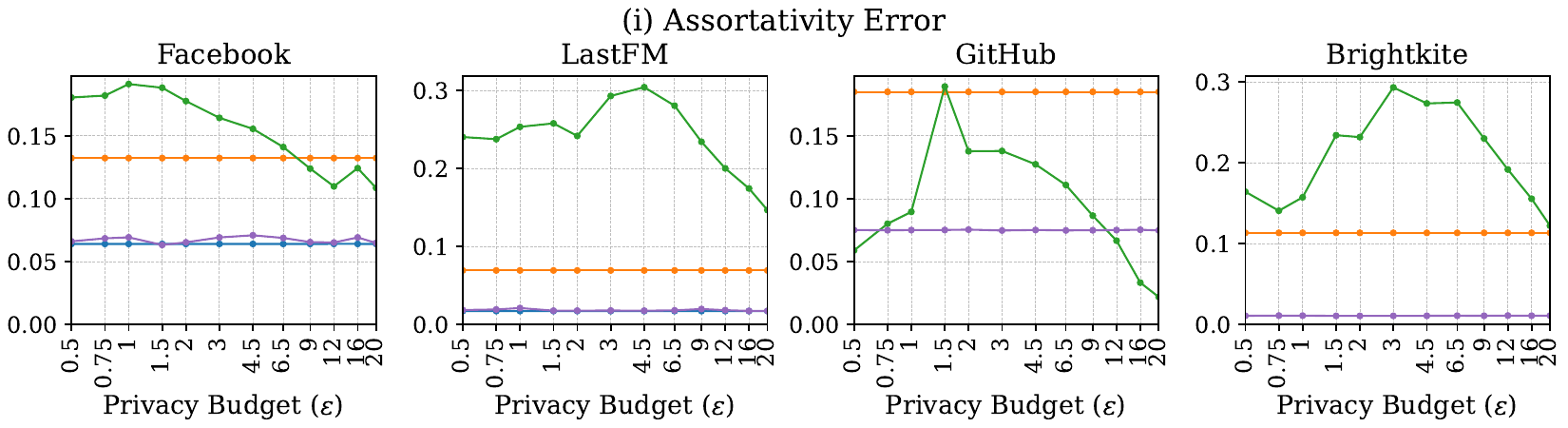}
    \hfill
    \includegraphics[width=0.498\linewidth]{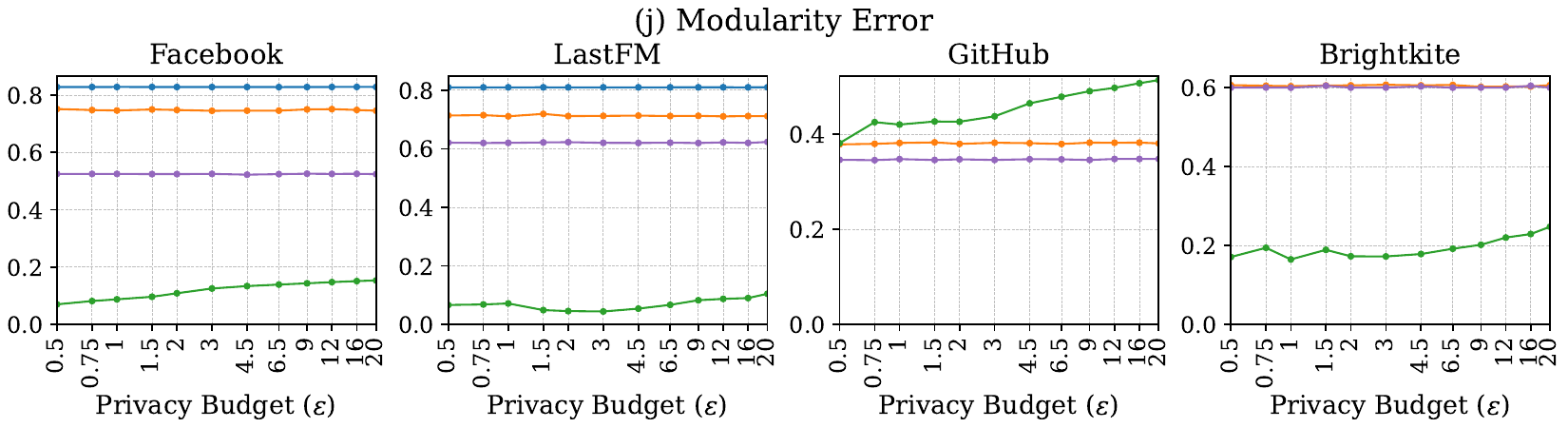}
    \includegraphics[width=0.25\linewidth]{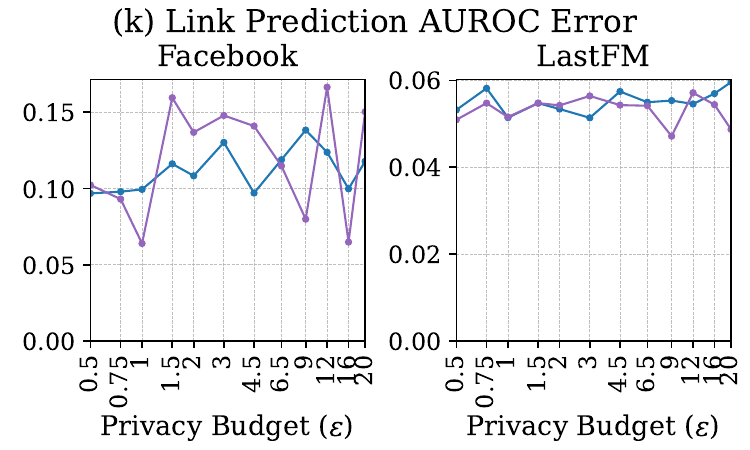}
    \hfill
    \includegraphics[width=0.145\linewidth]{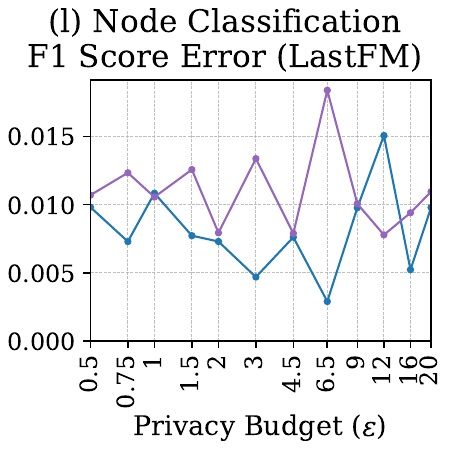}
    \hfill
    \raisebox{0.5cm}{\includegraphics[width=0.1\linewidth]{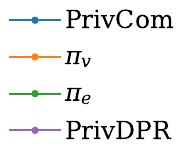}}
    \hfill
    \begin{minipage}[b]{0.472\linewidth}
        \vspace{-2mm}
        \caption{Results for Scenario 2. Distribution errors are quantified with the Wasserstein distance; the number of nodes, number of edges, and harmonic diameter with relative error; and remaining error-based metrics with absolute error. All values are averaged over 10 independent trials.}
        \label{fig:scenario-2-results}
    \end{minipage}
\end{figure*}

Results are presented in Fig.~\ref{fig:scenario-2-results}, with further plots available in our repository, similarly to Scenario~1.

PrivCom provides node-DP by approximating the Katz Index of the input graph, estimating the eigen-decomposition using a private variant of Oja's method, and then generating a noisy Katz matrix to construct a private graph via the graph Laplacian. Since the node set of the private output graph remains identical to that of the input graph, a large number of edges must be added to make the nodes indistinguishable, thereby ensuring node-DP. This explains the high error in the density and number of nodes for PrivCom.

Jian et al. recently proposed two node-DP algorithms: $\pi_v$, which relies on vertex perturbation, and $\pi_e$, which relies on edge perturbation. While promising results are claimed, our analysis raises concerns about their practical utility for real-world graphs.

At the core of $\pi_v$ lies two parameters, $0 < p < 1$ and $0 < q < 1$, which control the fraction of vertices to be removed and added (with randomly connected edges), respectively. For DP to hold, these parameters must satisfy the constraints: $1/p \le e^\epsilon$, and $p + (1-p)2^n / (1-q) \le e^\epsilon$, where $n = |V|$. However, since $0 < q < 1$, we get that $p + (1-p)2^n \le e^\epsilon$. Then, $p \ge 1 - (e^\epsilon - 1)/(2^n - 1)$, which means for $n > 1$, the constraint implies $p \ge 1 - e^\epsilon/2^n$, regardless of the choice of $q$. If $p \simeq 1$, then all nodes are removed, meaning the output graph is randomly generated, providing trivial privacy. For example, even with a generous privacy budget of $\epsilon=100$ and a relatively small graph of $n = 200$, we still find that $p \ge 1 - 1.673\times10^{-17}$, which converges ever closer to $1$ as $n$ increases.

The $\pi_e$ algorithm faces a similar issue. Although it introduces an optimal estimation procedure for its probability parameters, in practice, most edges are removed before random edges are added, severely undermining the utility.

As shown in Fig.~\ref{fig:scenario-2-results}, both $\pi_v$ and $\pi_e$ exhibit substantially larger errors across all metrics. To emphasise, this effect exists even in the simplest structural preservation metrics, such as errors in the number of nodes, the number of edges, and the density. A failure to preserve these foundational structural properties inevitably impacts the preservation of more complex, nuanced topological properties that capture cohesion and community structure, thus offering considerably less utility for downstream analytical tasks than other approaches.

In contrast, PrivDPR preserves most structural properties well, performing at least as well as other node-DP methods across most statistics. Its advantage likely stems from being the only deep-learning-based node-DP method that synthesises graphs via a Page-Rank-based approach.  However, this comes at a cost in scalability, as adjacency matrix parameters must be stored in GPU memory, which limits the dataset size. Nevertheless, PrivDPR establishes itself as a SotA method for node-DP graph synthesis, provided that scalability is not a constraint for practitioners.

\subsubsection{Empirical Privacy Results and Discussion}
We present the results for empirical privacy preservation in Table~\ref{tab:node-dp-deanonymisation-facebook}. Since the $\pi_v$ and $\pi_e$ methods produced graphs that were drastically different from their inputs, we evaluated attacks only on PrivCom and PrivDPR.

As in Scenario 1, results do not vary with privacy budget, likely for the same reasons mentioned in Sec.~\ref{evaluations.scenario-1.empirical-privacy}. Specifically, they again suggest that higher-order structures tend to not be significantly affected by DP noise. However, the results differ across methods: PrivCom offers weaker protection than PrivDPR, especially for edge correctness, where $50\%$ of edges can be correctly mapped, compared to $0.4\%$ for PrivDPR, indicating much stronger protection. PrivCom's tendency to produce overly dense graphs likely inflates edge correctness, as more edges increase the likelihood of overlap with the original graph.

Both methods perform reasonably well on the symmetric substructure score, which measures the number of conserved edges relative to the union of both graphs, with PrivDPR again offering slightly better protection.

Finally, and as mentioned briefly in Sec.~\ref{evaluations.scenario-2.o3}, evaluating node de-anonymisation attacks against node-DP mechanisms is challenging. Most such mechanisms do not maintain a mapping between the sensitive input and privatised output graphs, making ground-truth-based metrics infeasible. This necessitates reliance on indirect and ambiguous ground-truth–agnostic measures, which complicates the interpretation and contextualisation of results.

\begin{table}
    \small
    \centering
    \begin{tabular}{c|ccc|ccc}
        \hline \multirow{2}{*}{\textbf{Method}} & \multicolumn{3}{c|}{\multirow{2}{*}{\textbf{Edge Correctness} (\%)}} & \multicolumn{3}{c}{\textbf{Symmetric Substructure}} \\
        & & & & \multicolumn{3}{c}{\textbf{Score} (\%)} \\
        \hline & $\epsilon = 1$ & $\epsilon = 3$ & $\epsilon = 9$ & $\epsilon = 1$ & $\epsilon = 3$ & $\epsilon = 9$ \\
        \hline PrivCom & \num{50.081} & \num{50.058} & \num{51.152} & \num{1.067} & \num{1.067} & \num{1.090} \\
        \hline PrivDPR & \num{0.258} &  \num{0.348} & \num{0.246} & \num{0.221} & \num{0.299} & \num{0.211} \\
        \hline
    \end{tabular}
    \caption{Results of the de-anonymisation attack~\cite{Pedarsani13} against relevant node-DP methods, on the Facebook dataset.}
    \label{tab:node-dp-deanonymisation-facebook}
\end{table}

\section{Future Directions and Limitations}\label{limitations-future-work}
\paragraph{Future Directions}
We identify and summarise open avenues for future research. First, our systemisation revealed that temporal and attributed graphs have received limited attention. As datasets grow in scale and semantic richness, advancing privacy-preserving methods for these graph models becomes increasingly important.

Second, our systemisation highlighted that trust models beyond the centralised paradigm (e.g. local, shuffle, etc.) for graph data are underexamined. Further investigations on these models may enable applications in settings where such trust assumptions are inherent.

Third, our findings revealed a strong focus on edge-DP methods. However, in many cases, nodes represent entities that require privacy, making node-DP equally, if not more, critical. Yet, our evaluations showed that existing node-DP approaches offer insufficient utility for practical use, underscoring the need for improvement. 

Fourth, we observed an emerging trend of using deep learning techniques to generate private graphs, due to their ability to model complex structural and semantic patterns. While promising, this area remains underexplored, with current methods still in their infancy. Developing methods that balance privacy guarantees with the expressive power of deep learning models is a valuable direction.

Fifth, most existing methods use DP mechanisms that neglect dependencies. Since graphs inherently capture relationships, DP definitions that do not account for this cannot provably guarantee privacy. Accounting for dependencies inevitably requires more noise, likely degrading utility. Small-world networks, common in social graphs, offer a promising starting point due to their short-range dependencies~\cite{Watts98}. The alignment of BP with policy-graph structures represents a promising avenue for future research.

Sixth, we found few methods explicitly targeting the privacy of edge attributes (beyond edge-weights), while keeping structure public. While this privacy notion is strictly weaker than edge-DP, we posit that it may offer greater utility given the established trade-off. This is particularly valuable in domains where protecting edge attributes suffices, e.g., transactional or communication networks.

Last, our evaluation results highlighted ongoing challenges, particularly in preserving clustering and community structures, and the inherent tension between privacy and predictive utility when their targets align. Each presents valuable future directions.

\vspace{-0.825mm}
\paragraph{Limitations} Despite the contributions of this work, we acknowledge the following limitations. First, our scope was confined to indistinguishability-based notions of privacy, focusing on DP and its variants. Alternative notions of privacy such as $k$-anonymity~\cite{Samarati98}, $l$-diversity~\cite{Machanavajjhala07}, information-theoretic definitions~\cite{Bloch21}, multiparty computation~\cite{Lindell20}, or other cryptographic approaches, were not considered. Extending systematisation to these paradigms would be a valuable contribution.

Second, our illustrative evaluations benchmark only social network results. Whether these findings generalise to other domains remains an open question; future work could instantiate our framework across diverse application areas.

Third, as noted in Sec.~\ref{evaluations}, larger datasets were excluded due to computational inefficiencies in specific methods. While our selected datasets reveal general trends across varying sizes, scale-dependent effects would be more apparent on larger graphs. With many recent graph datasets~\cite{snap, ogb, nr} far exceeding our scale, the primary constraint is computational resources rather than data availability.

\section{Conclusion}\label{conclusion}
In this work, we proposed a novel systemisation of differentially private graph release methods, with a particular emphasis on critical practical considerations. The systemisation fuses a modular taxonomy of existing approaches with dimensions of vulnerabilities and attacks, and practitioner-defined contexts. These additions equip both experts and non-experts to select, apply, and evaluate suitable methods for real-world use cases. In doing so, we have contributed to improving the understandability of DP graph release approaches, mitigating the risks of invalid use and misleading claims about data protection. To demonstrate the use of our framework, we conducted two extensive empirical evaluations across social networks, providing a comprehensive benchmark within this domain. Finally, by releasing all code and results as open-source, we encourage reproducibility and further research.

\begin{acks}
    This research was supported by the Commonwealth through an Australian Government Research Training Program Scholarship\footnote{DOI: \href{https://doi.org/10.82133/C42F-K220}{10.82133/C42F-K220}}. The authors also thank UNSW and the anonymous reviewers.
\end{acks}
\bibliographystyle{ACM-Reference-Format}
\bibliography{references}

\appendix

\section{Formal Definitions}\label{formal-definitions}
\begin{table*}[t]
    \footnotesize
    \centering
    \renewcommand{\arraystretch}{1.35} %
    \begin{tabular}{r|cl}
        \hline \small \textbf{Metric} & \small \textbf{Formula} & \small \textbf{Additional Notes} \\
        \hline Number of Nodes & $|V|$ & \\
        \hline Number of Edges & $|E|$ & \\
        
        \hline \multirow{2}{*}{(Undirected) Density$(G)$} & \multirow{2}{*}{$\frac{|E|}{(|V|(|V|-1) / 2}$} & \\ \\
        
        \hline \multirow{2}{*}{Harmonic Diameter$(G)$} & \multirow{2}{*}{$\frac{|\text{SD}|}{\sum_{x \in \text{SD}} x^{-1}}$} & \multirow{2}{*}{SD is the set of shortest distances between all pairs of vertices.} \\ \\

        \hline \multirow{3}{*}{(Degree) Assortativity$(G)$} & \multirow{3}{*}{$\frac{\sum_{xy} xy(e_{xy} - a_x b_y)}{\sigma_a \sigma_b}$} & $e_{xy}$ is the proportion of edges joining nodes of degree $x$ and $y$, \\ & & $
        \{a, b\}_{*}$ are the proportions of edges incident on vertex $*$, \\ & & and $\sigma_a$, $\sigma_b$ are the standard deviations of $a_x$, $b_y$. \\
        
        \hline \multirow{3}{*}{Modularity$(G, c)$} & \multirow{3}{*}{$\frac{1}{2|E|} \sum_{i,j \in V \times V} \big( A_{ij} - \frac{\text{Deg.}(i) \text{Deg.}(j)}{2|E|} \big) \delta(c_i, c_j)$} & $\delta(c_i, c_j)$ is $1$ if nodes $i$ and $j$ have the same community, else $0$, \\ & & $A_{ij}$ is 1 if an edge exists between $i$ and $j$, else 0, \\ & & $c$ is a partition of communities, and Deg. abbreviates Degree. \\
        
        \hline Degree$(v)$ & $|\{(u, v) \in E\}|$ & Number of edges incident to a given node $v$. \\
        
        \hline \multirow{2}{*}{Clustering Coefficient$(v)$} & \multirow{2}{*}{$\frac{2 T(v)}{\text{Degree}(v)(\text{Degree}(v)-1)}$} & \multirow{2}{*}{$T(v)$ is the number of triangles through vertex $v$.} \\ \\
        \hline \multirow{2}{*}{Betweenness Centrality$(v)$} & \multirow{2}{*}{$\sum_{s, t \in V \times V} \frac{|\text{SP}(s, t | v)|}{|\text{SP}(s, t)|}$} & $|\text{SP}(s, t | v)|$ is the number of shortest paths from $s$ to $t$ via $v$, \\ & & $|\text{SP}(s, t)|$ is the number of shortest paths from $s$ to $t$. \\
        \hline \multirow{2}{*}{Closeness Centrality$(v)$} & \multirow{2}{*}{$\frac{|V|-1}{\sum_{u \in V, u \neq v} \text{SD}(u, v)}$} & \multirow{2}{*}{SD is the shortest distance function between $u$ and $v$.} \\ \\

        \hline \multirow{2}{*}{PageRank$(v)$} & \multirow{2}{*}{$\frac{1-d}{|V|} + d \sum_{(u, v) \in E} \frac{\text{PageRank}(u)}{| \{(u, w) \in E \} |} $} & \multirow{2}{*}{$d$ is the dampening factor, usually set to \num{0.85}.} \\ \\

        \hline \multirow{3}{*}{Adjusted Rand Index$(A, B)$} & \multirow{3}{*}{$\frac{
\sum_{ij}\binom{n_{ij}}{2}
- \big[ \sum_i\binom{a_i}{2}\sum_j\binom{b_j}{2} \big] \big/ \binom{n}{2}
}{
\frac{1}{2} \big[ \sum_i\binom{a_i}{2}+\sum_j\binom{b_j}{2} \big]
- \big[ \sum_i\binom{a_i}{2}\sum_j\binom{b_j}{2} \big] \big/ \binom{n}{2}
}$} & $a_i$ is the size of true label $i$ (representing a community), \\ & & $b_j$ is the size of predicted label $j$, and \\ & & $n_{ij}$ is the number of nodes with true label $i$ and predicted label $j$. \\

        \hline \multirow{2}{*}{Normalised Discounted Cumulative Gain$(A, B)$} & \multirow{2}{*}{$\frac{\text{{DCG}$(B)$}}{\text{DCG}(A)}$; where $\text{DCG}(X) = \sum_{i=1}^{|V|} \frac{\text{PageRank}(i)}{\log_2(i+1)}$} & \multirow{2}{*}{$A$ is the true ranking, $B$ is the predicted ranking.} \\ \\
        
        \hline Absolute Error & $|y - \hat{y}|$ & $y$ = true value; $\hat{y}$ = measured value. \\
        \hline Relative Error & $|y - \hat{y}| / y$ & $y$ = true value; $\hat{y}$ = measured value. \\

        \hline \multirow{2}{*}{$F_1$ Score} & \multirow{2}{*}{$\frac{2\text{TP}}{2\text{TP} + \text{FP} + \text{FN}}$} & \multirow{2}{*}{T = true; F = false; P = positive; N = negative.} \\ \\
        
        \hline \multirow{2}{*}{Edge Correctness$(G_1, G_2)$} & \multirow{2}{*}{$\frac{|f(E_1) \cap E_2|}{|E_1|}$} & \multirow{2}{*}{$f$ is the node mapping, and $G_1$ is the true graph.} \\ \\
        \hline \multirow{2}{*}{Symmetric Substructure Score$(G_1, G_2)$} & \multirow{2}{*}{$\frac{|f(E_1) \cap E_2|}{|E_1| + |E(G_2[f(V_1)])| + |f(E_1) \cap E_2|}$} & \multirow{2}{*}{$f$ is the node mapping, square brackets induce a subgraph.} \\ \\
        \hline \multirow{2}{*}{Reconstruction Relative Absolute Error$(G_1, G_2)$} & \multirow{2}{*}{$\frac{|| G_1 - G_2||_F}{||G_1||_F}$} & \multirow{2}{*}{$G_*$ are adjacency matrices. As defined in~\cite{Azogagh25}.} \\ \\
        \hline 
    \end{tabular}
    \caption{Formal definitions of metrics. Graphs are notated as $G = (V, E)$, and vertex iterators are typically denoted by $u, v, w$.}
    \label{tab:metrics-definitions}
\end{table*}

\subsection{Differential Privacies for Graphs}
We provide formal definitions of the two graph DP notions that dominate the literature, and that our work focuses on: edge-DP~\cite{Karwa11} and node-DP~\cite{Kasiviswanathan13a}. We then provide definitions for the less common notions mentioned, namely partition-DP~\cite{Task14} and attribute-DP~\cite{Blocki13}.

\paragraph{Edge Differential Privacy~\cite{Karwa11}} A randomised algorithm $A: \mathcal{G} \mapsto \mathcal{Y}$ is $\epsilon$\textit{-edge-differentially private} if for all neighbouring graphs $G = (V, E)$, $G' = (V, E') \in \mathcal{G}$ and for all possible outputs $Y \subseteq \mathcal{Y}$, 
\begin{equation}\label{eq:graph-dp}
    \text{Pr}[A(G) \in Y] \le e^\epsilon \text{Pr}[A(G') \in Y],
\end{equation}
where $G, G'$ are neighbouring graphs if $|E \triangle E'| = 1$. Intuitively, $G$ and $G'$ differ on exactly one edge by its addition or removal, however the node set remains the same.

\paragraph{Node Differential Privacy~\cite{Kasiviswanathan13a}} A randomised algorithm $A: \mathcal{G} \mapsto \mathcal{Y}$ is $\epsilon$\textit{-node-differentially private} if for all neighbouring graphs $G, G' \in \mathcal{G}$ and for all possible outputs $Y \subseteq \mathcal{Y}$, Eq.~\ref{eq:graph-dp} holds. $G = (V, E)$, $G' = (V', E')$ are neighbouring graphs if there exists some $V^* \subseteq V \cap V'$ such that\footnote{The notation $G[S]$ indicates an induced subgraph of $G$ given vertices $S \subseteq V$, i.e. $G[S] = (S, E \cap (S \times S))$.} $G[V^*] = G'[V^*]$ and $\max(|V|, |V'|) - |V^*| = 1$. Intuitively, $G$ and $G'$ differ on exactly one node by its addition or removal. If removed, all incident edges are implicitly removed; if added, the node is connected by any number of edges within $\llbracket 0, n \rrbracket$.

\paragraph{Partition Differential Privacy~\cite{Task14}} A partitioned graph is one consisting of $k > 1$ connected components, where $G = \{g_1, \ldots, g_k\}$ and all $g_i$ are disjoint subgraphs of $G$ such that $G = \bigcup_{i=1}^k \{g_i\}$. A randomised algorithm $A: \mathcal{G} \mapsto \mathcal{Y}$ is $\epsilon$\textit{-partition-differentially private} if for all neighbouring partitioned graphs $G, G' \in \mathcal{G}$ and for all possible outputs $Y \subseteq \mathcal{Y}$, Eq.~\ref{eq:graph-dp} holds, where $G, G'$ are neighbouring if $|G \triangle G'| = 1$.
Intuitively, $G$ and $G'$ differ on exactly one connected component by its addition or removal.

\paragraph{Vertex (and Edge) Attribute Differential Privacy~\cite{Blocki13}} A vertex-attributed graph $(G = (V, E), \ell)$ is a graph with a labelling function $\ell: V \mapsto \mathbb{R}^m$ that encodes information about its vertices. A randomised algorithm $A: \mathcal{H} \mapsto \mathcal{Y}$ is $\epsilon$\textit{-vertex-attribute-differentially private} if for all neighbouring attributed graphs $(G, \ell), (G, \ell') \in \mathcal{H}$ and for all possible outputs $Y \subseteq \mathcal{Y}$,
\begin{equation}\label{eq:attribute-dp}
    \text{Pr}[A((G, \ell)) \in Y] \le e^\epsilon \text{Pr}[A((G, \ell')) \in Y].
\end{equation}
Adapting the neighbouring notion from Blocki et al.~\cite{Blocki13} to privatise attributes \textit{only}, we define two vertex-attributed graphs $(G = (V, E), \ell), (G = (V, E), \ell')$ as neighbouring if there exists some $v \in V$ where $\ell(v) \neq \ell'(v)$ and for all $u \in V$ where $u \neq v$, $\ell(u) = \ell'(u)$. Intuitively, $(G, \ell)$ and $(G, \ell')$ differ on exactly one vertex attribute vector, while the structural information ($G$) remains the same. Note the current expression assumes vertex attributes, but is trivially adaptable to provide $\epsilon$\textit{-edge-attribute-differential privacy} instead.

\subsection{Metrics}
Formal definitions of all metrics used throughout the work are provided in Table~\ref{tab:metrics-definitions}.

\section{Systemisation, Annotated for Scenarios 1 \& 2}\label{appendix.systemisation-annotated}
\begin{figure*}[t]
    \centering
    \includegraphics[width=\textwidth]{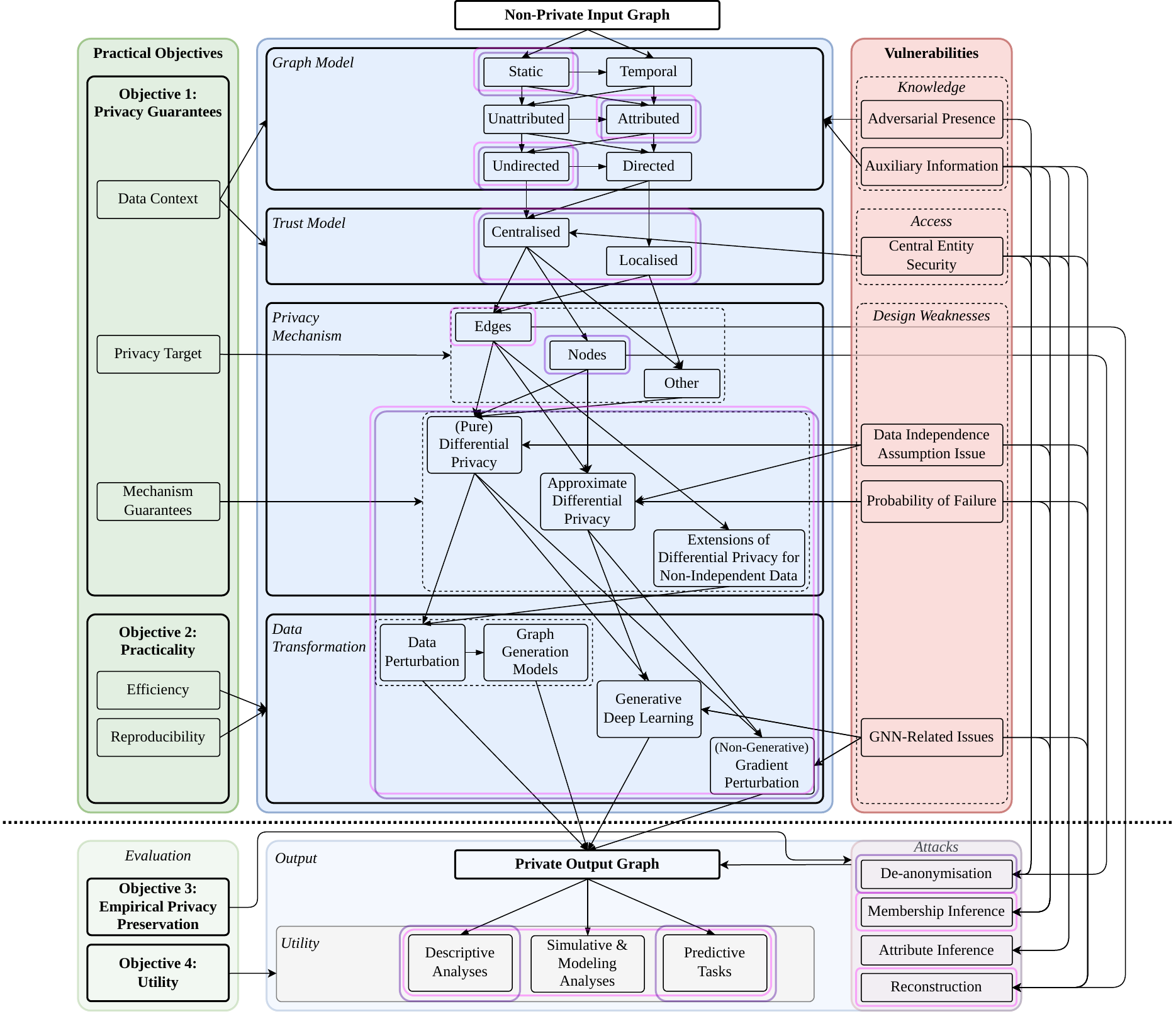}
    \caption{Our systemisation, annotated to illustrate the application of our systemisation to the two exemplar scenarios. The set of overlaid pink boxes represent modules relevant to the first scenario (Sec.~\ref{evaluations.scenario-1}); the set of overlaid purple boxes similarly relate to the second scenario (Sec.~\ref{evaluations.scenario-2}).}
    \label{systemisation-annotated-appendix}
\end{figure*}
For clarity, in Fig.~\ref{systemisation-annotated-appendix} we present an annotated version of our systemisation diagram to visually represent the application of our systemisation to each of the two exemplar scenarios explored in our work (Secs.~\ref{evaluations.scenario-1}, \ref{evaluations.scenario-2}).

\section{Additional Content}\label{appendix.further-details}
Our supplementary repository\textsuperscript{\ref{fn:repo}} contains all source code, numerical and tabular results, and additional plots omitted from this paper. Notably, these plots include CCDFs\footnote{Complementary cumulative distribution functions. This follows standard practice for analysing social graphs (the focus of our two scenarios), which are often assumed to follow power-law distributions~\cite{Barabasi16}.} for various distribution results and raw measurements plotted against dataset statistics. High-resolution renderings of all plots are also included.
\end{document}